\documentclass[12pt]{article}

\usepackage[usenames,dvipsnames,svgnames,table]{xcolor} 
\usepackage{kotex}
\usepackage{graphicx}
\usepackage{epsfig}
\usepackage{amssymb, amsmath, mathtools}
\usepackage{verbatim}
\usepackage{harvard} 
\usepackage[affil-it]{authblk}
\usepackage{mathrsfs}
\usepackage{here}
\usepackage{makecell}
\usepackage{tikz}
\usepackage{enumerate}
\usepackage[shortlabels]{enumitem}
\usepackage{yfonts}
\usepackage{comment}
\usepackage{longtable}
\usepackage{multirow} 
\usepackage{algorithm,algpseudocode}
\usepackage{subcaption} 
\usepackage{arydshln} 

\usepackage{epstopdf}
\AppendGraphicsExtensions{.pdf}

\usepackage{hyperref}
\usepackage[colorinlistoftodos, textsize=scriptsize]{todonotes} 

\usepackage[framemethod=TikZ]{mdframed}
\mdfdefinestyle{JL}{%
    linecolor=blue,
    outerlinewidth=1pt,
    roundcorner=2pt,
    innertopmargin=0.1\baselineskip,
    innerbottommargin=0.1\baselineskip,
    innerrightmargin=2pt,
    innerleftmargin=2pt,
    backgroundcolor=orange!100!white}
\newenvironment{JL}{\begin{mdframed}[style=JL]  \footnotesize} { \end{mdframed}}   
\newcommand{\bJ}{\begin{JL}}
\newcommand{\eJ}{\end{JL}}

\includecomment{note} 

\newtheorem{theorem}{Theorem}[section]
\newtheorem{lemma}[theorem]{Lemma}
\newtheorem{definition}[theorem]{Definition}
\newtheorem{proposition}[theorem]{Proposition}
\newtheorem{corollary}[theorem]{Corollary}

\newtheorem{remark}[theorem]{Remark}

\everymath{\displaystyle}

\newcommand{\half}{\frac{1}{2}}

\newcommand{\ra}{\rightarrow}

\newcommand{\be}{\begin{equation}}
\newcommand{\ee}{\end{equation}}
\newcommand{\bea}{\begin{eqnarray*}}
\newcommand{\eea}{\end{eqnarray*}}
\newcommand{\bean}{\begin{eqnarray}}
\newcommand{\eean}{\end{eqnarray}}
\newcommand{\ben}{\begin{enumerate}}
\newcommand{\een}{\end{enumerate}}
\newcommand{\bi}{\begin{itemize}}
\newcommand{\ei}{\end{itemize}}
\newcommand{\brem}{\begin{remark}}
\newcommand{\erem}{\end{remark}}
\newcommand{\bcen}{\begin{center}}
\newcommand{\ecen}{\end{center}}
\newcommand{\bsv}{\begin{semiverbatim}}
\newcommand{\esv}{\end{semiverbatim}}

\newcommand{\bt}{\begin{theorem}}
\newcommand{\et}{\end{theorem}}
\newcommand{\bl}{\begin{lemma}}
\newcommand{\el}{\end{lemma}}
\newcommand{\bd}{\begin{definition}}
\newcommand{\ed}{\end{definition}}
\newcommand{\bc}{\begin{corollary}}
\newcommand{\ec}{\end{corollary}}
\newcommand{\bp}{\begin{proposition}}
\newcommand{\ep}{\end{proposition}}

\newcommand{\bfd}{\mathbf{d}}
\newcommand{\bff}{ \mathbf{f}}
\newcommand{\bfm}{ \mathbf{m}}

\newcommand{\bfs}{ \mathbf{s}}

\newcommand{\bfu}{ \mathbf{u}}

\newcommand{\bfx}{ \mathbf{x}}
\newcommand{\bfy}{ \mathbf{y}}

\newcommand{\bfY}{ \mathbf{Y}}  
\newcommand{\bfZ}{\mathbf{Z}}

\newcommand{\bfone}{{\bf 1}}

\newcommand{\bbE}{{ \mathbb{E}}}

\newcommand{\bbR}{ \mathbb{R}}

\newcommand{\calQ}{\mathcal{Q}}

\newcommand{\bmu}{ \boldsymbol{\mu}}

\newcommand{\btheta}{ \boldsymbol{\theta}}

\newcommand{\bed}{\begin{itemize}}
\newcommand{\eed}{\end{itemize}}

\newcommand{\bfg}{ \mathbf{g}}
\newcommand{\bfp}{ \mathbf{p}}
\newcommand{\bfI}{ \mathbf{I}}
\newcommand{\bfJ}{ \mathbf{J}}
\newcommand{\bfV}{ \mathbf{V}}
\newcommand{\bvarepsilon}{ \boldsymbol{\varepsilon}}
\newcommand{\bfeta}{ \boldsymbol{\eta}}
\newcommand{\bsigma}{ \boldsymbol{\sigma}}
\newcommand{\bzero}{\boldsymbol{0}}
\newcommand{\bone}{\boldsymbol{1}}

\graphicspath{{fig/}} 

\title{Variational Bayes method for ordinary differential equation models}
\author{Hyunjoo Yang and Jaeyong Lee}
\affil{Department of Statistics \\ Seoul National University}

\begin{document}

\maketitle

\begin{abstract}
Ordinary differential equations (ODEs) are a mathematical model used in many application areas such as climatology, bioinformatics, and chemical engineering with its intuitive appeal to modeling. Despite ODE's wide usage in modeling, the frequent absence of their analytic solutions makes it challenging to estimate ODE parameters from the data, especially when the model has lots of variables and parameters. This paper proposes a Bayesian ODE parameter estimating algorithm which is fast and accurate even for models with many parameters. The proposed method approximates an ODE model with a state-space model based on equations of a numeric solver. It allows fast estimation by avoiding computations of a complete numerical solution in the likelihood. The posterior is obtained by a variational Bayes method, more specifically, the \textit{approximate Riemannian conjugate gradient method} \cite**{honkela2010approximate}, which avoids samplings based on Markov chain Monte Carlo (MCMC). In simulation studies, we compared the speed and performance of the proposed method with existing methods. The proposed method showed the best performance in the reproduction of the true ODE curve with strong stability as well as the fastest computation, especially in a large model with more than 30 parameters. As a real-world data application, a SIR model with time-varying parameters was fitted to the COVID-19 data. Taking advantage of the proposed algorithm, more than 50 parameters were adequately estimated for each country. 

\end{abstract}


\section{Introduction} \label{s:intro}
Ordinary differential equations (ODEs) are a basic mathematical modeling tool describing dynamical systems. By describing local changes, the ODE models the global system. With strong interpretability, ODEs are used in many application areas such as climatology, bioinformatics, disease modeling, and chemistry. 

Despite the usefulness of modeling, there are difficulties in estimating their parameters from the data. In most cases, the solution of ODE cannot be calculated analytically. For an ODE model (ODEM), a nonlinear regression model whose mean function is a solution of ODE, repeated computations of numerical solutions in the likelihood make the inference very slow. Aside from the speed, an accurate estimate itself is challenging because the ODE solutions as regression curves make the likelihood surface extremely messy.

From the frequentist side, the two-step approach was proposed by \citeasnoun{swartz1975discussion} and \citeasnoun{varah82spline}. In the first step,  the observations are fitted with a function estimation method without considering the ODE. In the second step, the parameters of ODE are estimated by minimizing the difference between the estimated function and the ODE.

\citeasnoun{ramsay2005functional} proposed the penalized likelihood approach, which penalizes deviation from the differential equation. The penalized likelihood is optimized iteratively with respect to the parameters of the differential equation and the regression function. \citeasnoun**{ramsay2007parameter} proposed the generalized profiling method whose computation consists of outer and inner loops. In the outer loop, the parameters of the differential equation are optimized for the likelihood function. Whenever the regression function is required in the outer loop, it is obtained by minimizing the penalized likelihood called the inner loop. 

\citeasnoun{liang2008parameter} and \citeasnoun**{liang2010estimation} proposed two-step approaches where the local polynomial regression is adopted in the observation fitting stage. \citeasnoun{hall2014quick} proposed a one-step estimation method in which the tuning parameter and the parameters in the ODE are estimated simultaneously by minimizing an objective function. 

\citeasnoun**{gelman1996physiological} and \citeasnoun**{huang2006hierarchical} fitted the nonlinear regression model for the Bayesian pharmacokinetic model and hierarchical HIV dynamic model, respectively, and Markov chain Monte Carlo (MCMC) samples were generated with a numeric solver of ODE. \citeasnoun{campbell2012smooth} proposed smooth functional tempering, a population MCMC method with differing tempering parameters. \citeasnoun**{Bhaumik2015bayesian} proposed a Bayesian version of the two-step approach and proved the Bernstein-von Mises theorem holds.  \citeasnoun**{dass2017laplace} proposed a combination of Laplace approximation and grid sampling method. \citeasnoun**{lee2018inference} used a particle filter method for the relaxed state-space model.
For nonparametric Bayesian modeling, a gradient matching method using Gaussian Processes (GPs) as a data regression model was proposed by \citeasnoun**{calderhead2008accelerating}.  This methodology was later evolved into the adaptive gradient matching (AGM) of \citeasnoun**{dondelinger2013ODE} and the GP-ODE generative model of \citeasnoun{wang2014gaussian}.
 
Despite these various methods, there are still limitations in that performance comparisons have been made only for small ODE models and that even for these small models, they occasionally show a poor inference. This paper proposes a Bayesian ODE parameter estimating algorithm that is fast and accurate for even somewhat large models with lots of parameters. The algorithm exploits a state-space model and a variational Bayes approximation. Following \citeasnoun{lee2018inference}, ODE models are relaxed to a state-space model based on equations of a numeric solver such as the Runge-Kutta method. It allows fast estimation by avoiding computations of a complete numerical solution in a likelihood. The posterior is obtained by the variational Bayes method. By employing the approximate Riemannian conjugate gradient method of \citeasnoun**{honkela2010approximate}, which perfectly matches with our model, the estimators are obtained fast. In addition to high speed, the proposed method showed strong stability even in a large model with more than 30 parameters in simulation studies, while  all the other competing estimators have not provided valid inferences for the same large model.  

The rest of the paper is organized as follows. Section \ref{s:method} describes the proposed method in detail, along with the relaxed state-space model and some background of variational Bayes. Simulation experiments comparing the speed and performance of our algorithm with others are provided in Section \ref{s:simul}.  In Section \ref{s:appl}, as an application to real-world data, a time-varying SIR model, which replaced the parameters of the simple SIR model with time-varying parameters using B-spline bases, was fitted to the COVID-19 data. We end the paper with a discussion in Section \ref{s:discuss}. Some details for the algorithm implementation and tuning parameters are given in Appendix.

\section{Method} \label{s:method}
\subsection{Ordinary Differential Equation and State-Space Model}
An ordinary differential equation represents changes of variables with their derivatives and themselves. Let $p$-dimensional functions $\bfx(t)$ satisfy an ODE 
\begin{align*}
\dot{\bfx}(t)=\bff(\bfx(t),t\ ;\btheta),\qquad t\in[0,T],
\end{align*} 
with the ODE parameters $\btheta \in \Theta \subset \bbR^q$. Observations are modelled with  errors by
\begin{align}
\label{ODE model}
\begin{split}
\bfy_i=&\ \bfx_i+\bvarepsilon_i,\quad\bvarepsilon_i\overset{iid}{\sim}\text{N}(\bzero,\lambda^{-1}\bfI_p), \\
\dot{\bfx}(t)&=\bff(\bfx(t),t\ ;\btheta).
\end{split}
\end{align} 
Here, $ \bfx_i:=\bfx(t_i) $ for $\ i=0,1,\dots,n $ and $ 0\leq t_0\leq t_1\leq\cdots\leq t_n\leq T$. Given these data, our goal is to estimate the ODE parameters $\btheta$  quickly and accurately. Further, since an ODE solution is determined by initial states $\bfx_0$ as well as $\btheta$, the estimates of $\bfx_0$ can be as important as those of $\btheta$.

In many cases, ODEs have no analytic solutions, and numerically computed solutions are exploited instead. Higher accuracy of numerical solutions requires a smaller step size, which implies a higher cost of computation. Repeated computations of the numerical solution in the likelihood become a significant cause for a slow inference of ODE models. 

As a strategy to avoid these difficulties, we follow \citeasnoun{lee2018inference} and approximate the ODE with a state-space model. The original ODE model (\ref{ODE model}) can be relaxed to the following state-space model:
\begin{align}
\label{SSM}
\begin{split}
\bfy_i&=\bfx_i+\bvarepsilon_i,\quad \bvarepsilon_i\overset{iid}{\sim}\text{N}(\bzero,\lambda^{-1}\bfI_p), \quad i=0,1,\dots,n,\\
\bfx_{i+1}=&\ \bfg(\bfx_i,t_i,\btheta)+\bfeta_i,\quad \bfeta_i\overset{iid}{\sim}\text{N}(\bzero,\tau\bfI_p), \quad i=0,1,\dots,n-1.
\end{split}
\end{align}

In the  \textit{ODE state-space model} (OSSM) (\ref{SSM}), $ \bfg(\cdot) $ is an approximating function based on a numeric method. We choose the 4th-order Runge-Kutta method throughout this paper, which is one of the most widely used numeric solvers of ODEs. The function $g$ is expressed as 
\begin{align}
\label{Runge}
\begin{split}
\bfg(\bfx_i^*,t_i,\btheta)=&\ \bfx_i^*+\frac{1}{6}(K_{i1}+2K_{i2}+2K_{i3}+K_{i4}),\\
K_{i1}=&\ h_{i+1}\cdot \bff(\bfx_i^*,t_i;\btheta),\\
K_{i2}=&\ h_{i+1}\cdot \bff(\bfx_i^*+\frac{1}{2}K_{i1},t_i+\frac{1}{2}h_{i+1};\btheta),\\
K_{i3}=&\ h_{i+1}\cdot \bff(\bfx_i^*+\frac{1}{2}K_{i2},t_i+\frac{1}{2}h_{i+1};\btheta),\\
K_{i4}=&\ h_{i+1}\cdot \bff(\bfx_i^*+K_{i3},t_i+h_{i+1};\btheta),
\end{split}
\end{align}
where $h_{i+1} = t_{i+1} - t_i$ for $i = 0,1, \ldots, n-1$. 
The computation of $\bfg(\cdot)$  is much simpler than that of the whole numerical solution because it needs just one-step numerical calculations based on each interval $h_i$ from the observation time points. If a time interval $ h_i$ is so large as to threaten the stability of the approximation severely, we can divide each interval into $ m $ subintervals and apply the numerical method over the $ m $ subintervals repeatedly. Let $m$ be called the step size.

The approximated model requires an error term $\bfeta_i$.  Here $\tau$, the variance of the errors,  is a constant tuning parameter that determines the amount of allowable uncertainty. When the selected numerical method has reliable performance, the smaller $\tau$ means that the states $\bfx$'s should follow the ODE solution more closely. As $m\rightarrow\infty$  and $\tau\rightarrow0$, the OSSM approaches to the true ODE model ($\ref{ODE model}$). In fact, \citeasnoun{lee2018inference} proved that under some regularity conditions, the posterior of OSSM (\ref{SSM}) converges to that of the true ODE model ($\ref{ODE model}$). For details, see \citeasnoun{lee2018inference}.
 
The parameters to be estimated are the ODE parameters $\btheta$ and initial states $\bfx_0$, which determine the ODE solution. The extent of the measurement noise, $\lambda$, and the subsequent states $\bfx_i$ in times $t_1,\dots,t_n$ are additionally inferred. Recall that the tuning parameter $\tau$ is a constant.

The priors for the parameters in this paper are as follows: 
\begin{align}
\label{prior}
\begin{split}
\lambda\sim& \ \text{Gamma}(A_0,B_0), \\
\theta_k\sim& \ \text{Unif}(a_{0k},b_{0k}), \ \text{for }k=1,\dots,q, \\
x_{0j}\sim& \ \text{Unif}(c_{0j},d_{0j}),\ \text{for }j=1,\dots,p. 
\end{split}
\end{align}
The gamma prior for $\lambda$ has  mean $A_0/B_0$ and variance $A_0/B_0^2$.   The priors for ODE parameters and initial states are uniform distributions that are fully independent of each other. The supports $ (a_{0k},b_{0k}) $ of the  $\btheta$ priors are chosen based on scientific knowledge and sufficiently large to contain the true parameter values. 
For $\bfx_0$, some reasonable bounds $ (c_{0j},d_{0j}) $ can be considered from the observed data.

\subsection{Posterior Inference with Variational Bayes Approximation}\label{subsec:VB}
The variational inference or variational Bayes is a posterior computation method that approximates a posterior density function using the Kullback-Leibler divergence. First appearing in  \citeasnoun**{jordan1999introduction}, it has gained considerable popularity as a posterior computation method due to the high speed of computation.  

The variational Bayes predefines a specific family of densities. Based on the Kullback-Leibler divergence, the member in that family closest to the true posterior is regarded as the approximate posterior. When $\btheta$ represents a vector including both unknown parameters and latent variables, the joint density of $\btheta$ and observations $\bfy$ is 
$$ p(\btheta,\bfy)=p(\btheta)p(\bfy|\btheta). $$ 
The posterior $p(\btheta|\bfy)$, often not easy to calculate, is approximated by a member of the predefined family $\mathcal{Q}$ using the Kullback-Leibler divergence (KL) as a measure of `closeness'. The density with the minimum KL,    $$  q^*(\btheta)=\underset{q(\btheta)\in\mathcal{Q}}{\arg \min} D_{KL}(q(\btheta)||p(\btheta|\bfy))=\underset{q(\btheta)\in\mathcal{Q}}{\arg \min}\int_{\btheta}q(\btheta)\log\frac{q(\btheta)}{p(\btheta|\bfy)}d\btheta $$ is selected as the approximate posterior. In practice, the cost function used is not  $D_{KL}$ but a variant 
\begin{align}
	C_{KL}=D_{KL}(q||p)-\log p(\bfy)= \int_{\btheta}q(\btheta)\log\frac{q(\btheta)}{p(\btheta,\bfy)}d\btheta
	\label{C}
\end{align}
that is more tractable and provides the equivalent result of minimization. For more details about the variational method, see \citeasnoun**{blei2017variational}. 

In the popular {\it mean-field approximation}, $\calQ$ consists of distributions whose variables are all mutually independent. 
This setting makes the calculation of the objective function ($\ref{C}$) relatively easy. With such an advantage, we also take this family for the posterior of $ (\lambda,\btheta,\bfx) $ in OSSM, having the form: 
\begin{align}
\label{vardist}
\begin{split} 
q(\lambda,\btheta,\bfx)&=\ q(\lambda)q(\btheta)q(\bfx),\\
q(\lambda)=&\ \text{Gamma}(A_\lambda,B_\lambda),\\
q(\btheta)=&\ \prod_{k=1}^q \text{TN}_{(a_{0k,},b_{0k})}(\theta_k;\ \mu_k,\sigma_k^2),\\
q(\bfx) =& \ \text{TN}_{(c_0,d_0)}(\bfx_0;\ \bfm_0,\bfV_0)\prod_{i=1}^n\text{N}(\bfx_i;\ \bfm_i,\bfV_i),\\ 
&\bfV_i =\ \text{diag}\{V_{i1},\dots,V_{ip}\}.
\end{split}
\end{align} 
The ODE parameters $\btheta$ and the latent variables $\bfx$ follow normal distributions as  marginal distributions. To be exact, for $\btheta$ and $\bfx_0$, which have the uniform priors in (\ref{prior}), truncated normal distributions with the supports of (\ref{prior}) would be more rigorous.  However, because the effect of tails in a normal distribution is generally negligible under small variances,  subsequent calculations treat them as normal distributions. The ranges $(a_{k0},b_{k0})$, $ (c_{0j},d_{0j}) $ in the uniform priors may be used as the bounds for the optimization procedure of the variational Bayes. 

Now the cost function of formula (\ref{C}) can be calculated with the priors (\ref{prior}) and the family $\mathcal{Q}$ (\ref{vardist}),  as follows:
\begin{align*}
C_{KL}=&\ \left(A_\lambda-A_0-\frac{p(n+1)}{2}\right)\psi(A_\lambda)+\left(A_0+\frac{p(n+1)}{2}\right)\log B_\lambda\\
&-A_\lambda-\log\Gamma(A_\lambda) -\frac{1}{2} \sum_{k=1}^{q}\log\sigma_k^2-\frac{1}{2}\sum_{i=0}^{n}\sum_{j=1}^{p}\log V_{ij}\\
&+\frac{1}{2\tau}\sum_{i=1}^{n}\bbE_q\lVert\bfm_i-\bfg(\bfx_{i-1},t_{i-1},\btheta)\rVert^2 +\frac{1}{2\tau}\sum_{i=1}^{n}\sum_{j=1}^{p}V_{ij}\\
&+\frac{A_\lambda}{B_\lambda}\left[B_0+\frac{1}{2}\sum_{i=0}^{n}\sum_{j=1}^{p}\left[(m_{ij}-y_{ij})^2+V_{ij}\right]\right] \\ 
&+ \text{constant terms}.
\end{align*}
$ C_{KL} $ is a function of the variational parameters $ (A_\lambda, B_\lambda, \bmu,\bsigma^2, \bfm, \bfV) $.  
Letting the partial derivatives  of $C_{KL}$  with respect to $A_\lambda$ and $B_\lambda$ equal to zero gives the following equations:
$$ A_\lambda = A_0+\frac{p(n+1)}{2}\ ,$$
$$ B_\lambda =  B_\lambda(\bfm,\bfV) = B_0+\frac{1}{2}\sum_{i=0}^{n}\sum_{j=1}^{p}\left[(m_{ij}-y_{ij})^2+V_{ij}\right].$$
By substituting the above expressions, we can eliminate $A_\lambda$ and $B_\lambda$ from $C_{KL}$ and obtain 
\begin{align}
\label{cost1}
\begin{split}
C_{KL}(\bmu,\bsigma^2, \bfm, \bfV) =&\ A_\lambda\log B_\lambda(\bfm,\bfV)+\frac{1}{2\tau}\sum_{i=1}^{n}\sum_{j=1}^{p}V_{ij} \\
&  -\frac{1}{2} \sum_{k=1}^{q}\log\sigma_k^2-\frac{1}{2}\sum_{i=0}^{n}\sum_{j=1}^{p}\log V_{ij} \\
&+\frac{1}{2\tau}\sum_{i=1}^{n}\bbE_q\lVert\bfm_i-\bfg(\bfx_{i-1},t_{i-1},\btheta)\rVert^2  \\
&+ \text{constant terms}.
\end{split}
\end{align}

The last expectation term in (\ref{cost1}) is nearly impossible to find its explicit form because of the nested structure of the function $\bfg(\cdot)$. To compute the value and express the term as a function of $ (\bmu,\bsigma^2, \bfm, \bfV) $, we use the Monte Carlo method.
When $M$ is the number of Monte Carlo samples and $\{ \btheta^{(s)},\bfx_0^{(s)} ,\bfx_1^{(s)},\cdots,\bfx_{n}^{(s)}\}_{s=1}^{M} $ are the Monte Carlo samples from $q(\cdot)$, 
\begin{equation}\label{monte}
	\begin{split}
		\bbE_q\lVert\bfm_i-\bfg(\bfx_{i-1},t_{i-1},\btheta)\rVert^2
		\approx&\ \frac{1}{M}\sum_{s=1}^{M}\left\lVert\bfm_i-\bfg(\bfx_{i-1}^{(s)},t_{i-1},\btheta^{(s)})\right\rVert^2.
	\end{split}
\end{equation}
Again, since the $q(\cdot)$ takes the mean-field assumption and $(\btheta^{(s)},\ \bfx^{(s)})$ follow the normal distributions with the parameters $ (\bmu,\bsigma^2, \bfm, \bfV) $, we can also get the samples in the following ways:
$$  \{\btheta^{(s)},\ \bfx^{(s)}\}_{s=1}^{M} = \{  \sqrt{\bsigma^2}\odot\bfZ_{\btheta}^{(s)}+\bmu,\ \sqrt{\bfV}\odot\bfZ_{\bfx}^{(s)}+\bfm \}_{s=1}^{M},   $$ 
where $\bfZ^{(s)} = (\bfZ_{\bfx}^{(s)}, \bfZ_{\btheta}^{(s)}) $ are samples from N($\bzero, \bfI$) and   the $\odot$ symbol denotes elementwise multiplication. The covariance  $\bfV$ in the last equation is regarded as a vector of its diagonal elements rather than a matrix. Finally, the cost function has the form:
\begin{align}
	\begin{split}
		C_{KL}(&\bmu,\bsigma^2, \bfm, \bfV)\\ =&\ A_\lambda\log B_\lambda(\bfm,\bfV) +\frac{1}{2\tau}\sum_{i=1}^{n}\sum_{j=1}^{p}V_{ij} -\frac{1}{2} \sum_{k=1}^{q}\log\sigma_k^2-\frac{1}{2}\sum_{i=0}^{n}\sum_{j=1}^{p}\log V_{ij}\\
		&+\frac{1}{2\tau}\frac{1}{M}\sum_{i=1}^{n}\sum_{s=1}^{M}\left\lVert\bfm_i-\bfg(\sqrt{\bfV_{i-1}}\odot\bfZ_{\bfx_{i-1}}^{(s)}+\bfm_{i-1},\ t_{i-1},\sqrt{\bsigma^2}\odot\bfZ_{\btheta}^{(s)}+\bmu)\right\rVert^2 \\
		&+\text{constant terms}.
	\end{split}
\end{align}

Since the cost function should be a function of only the variational parameters $ (\bmu,\bsigma^2, \bfm, \bfV) $, the samples $\bfZ^{(s)} $ must be provided before starting an optimization algorithm. To make a good Monte Carlo approximation with a small sample size $M$, we use a kind of quasi-Monte Carlo method. As is well known, when $F_Z(\cdot)$ indicates the cumulative distribution function (CDF) of the standard normal distribution, 
$$	F_Z^{-1}(U)\ \overset{iid}{\sim} \ \text{N}(0, 1),\quad \text{ for }\quad U\ \overset{iid}{\sim} \ \text{Unif}(0,1).$$ 
The idea is to use the most plausible samples from the uniform distribution to get balanced $Z$ samples. We can set $ \{z^{(s)}\}_{s=1}^{M} $ to the followings:
\begin{align}
\label{z}
\begin{split}
z^{(s)} = F_Z^{-1}(u^{(s)}),  \quad  \text{where}\quad u^{(s)} = \frac{s}{M} -\frac{1}{2M}\ \text{ for }\ s=1,\cdots, M.
\end{split}
\end{align} 
Especially if $M$ is an odd number, the center of the standard normal distribution $ 0 = F_Z^{-1}(0.5) $ corresponding to $s=(M+1)/2$, would be included in the samples so that the samples look ``plausible'' as schemed to be symmetric about the center and  reasonably well distributed.  Once we obtain a sample set values $ \{z^{(s)}\}_{s=1}^{M} $ according to (\ref{z}) the order of the index $s$ is randomly shuffled for each variables $\bfZ_{ \btheta}^{(s)},\ \bfZ_{\bfx_0}^{(s)} ,\ \bfZ_{\bfx_1}^{(s)},\ \cdots,\ \bfZ_{\bfx_{n}}^{(s)} $ to meet the mutual independence condition (mean-field). 

These schemes actually showed good results with a small sample size $M$ in the simulation studies. Under the simulation settings of Section \ref{s:simul}, the approximation of equation (\ref{monte}) was investigated. When $M=11$, they were approximately 90\% of the true values (that calculated with a large $M$). In other words, there was almost no difference in the estimation result between $M=11$ and larger $M$. Therefore, in all subsequent implementations, we set $M=11$.

The variational parameters are optimized using the \textit{approximate Riemannian conjugate gradient learning} of \citeasnoun**{honkela2010approximate}. This algorithm for fixed-form variational Bayes optimizes the cost function using a conjugate gradient method that exploits the Riemannian geometry of the space of the variational parameters. They applied the algorithm to a nonlinear state-space model, which belongs to the non-conjugate exponential family, and reported that their algorithm outperforms the existing gradient-based algorithms. For details, we refer to \citeasnoun**{honkela2010approximate}. The resulting algorithm for our model is presented in Appendix \ref{algorithm}, and the Jacobian matrix computations required in the algorithm are derived in Appendix \ref{Jacobian}.

If numerical problems occur during the optimization procedure, such as failure in the line search, the optimization process is completely restarted at a new starting point based on the prior, automatically. The algorithm's run time surely includes all these restart processes and nevertheless, it showed the fastest inference speed in the simulation study.

\section{Simulation study } \label{s:simul}
\subsection{Competitors}
A total of five estimators, including the proposed method, are compared in the simulation study.
The four methods to be compared with ours are as follows.
\bi
\item \textbf{Parameter cascade method (PC)} \\
The parameter cascade method by \citeasnoun**{ramsay2007parameter} is a representative of the frequentist approach in the ODE parameter inference. The method estimates parameters with nested optimization:
\begin{align*}
\hat\theta =\ & \underset{\theta}{\arg min}\sum_{i=0}^{n}\sum_{j=1}^{p}(Y_{ij}-\hat{x}_{ij}(\theta))^2 \quad \text{(outer)}, \\
\hat{\mathbf{x}}(\theta) =\ & \underset{\mathbf{x}\ \in\ B^p}{\arg min}\sum_{i=0}^{n}\sum_{j=1}^{p}(Y_{ij}-x_{j}(t_i))^2 + \lambda\sum_{j=1}^{p}\int[\dot{x}_j(t) - f_j(\mathbf{x}(t);t,\theta)]^2 dt  \quad \text{(inner),}
\end{align*} 
where B is a B-spline basis expansion with a degree of $D$ having $m$ knots at $ u_0\leq\cdots\leq u_{m-1} $,
$$ B(t)=  \sum_{j=0}^{m-D+2}c_{j}B_{j,D}(t), \ \ t\in[u_D,u_{m-D-1}] $$
when
\begin{align*}
    &B_{j,0}(t):=\Bigg\{
    \begin{array}{lr}
    1,&\quad u_j\leq t < u_{j+1}\\
    0,& \quad\;\;\: \text{otherwise}
    \end{array}
     ,\quad j = 0,\dots,m-2, \\
     &\text{and for }d>0, \\
	&B_{j,d}(t) :=\left(\frac{t-u_j}{u_{j+d}-u_j}\right) B_{j,d-1}(t)+\left(\frac{u_{j+d+1}-t}{u_{j+d+1}-u_{j+1}}\right)B_{j+1,d-1}(t), \quad j=0,\dots, m-d-2.
\end{align*}

To put it simply, it is a kind of least squares regression based on B-spline basis functions, but whose basis coefficients remain loyal to the ODE as well as the observations. They avoid the numeric ODE solver by using $\bff(\cdot)$ of the ODE for the measure of loyalty. In all of our experiments, the smoothing parameter $\lambda$ was chosen by the \textit{forwards prediction error} (FPE) of  \citeasnoun{ellner2007commentary}. The \textbf{CollocInfer}  package in \textbf{R} was used for implementation. The functions in the package implement their optimization with the `nlminb' function using C/C++.

\item \textbf{Delayed rejection \& adaptive Metropolis algorithm (DRAM)} \\
The delayed rejection \& adaptive Metropolis algorithm by \citeasnoun**{haario2006dram} was chosen as a general MCMC method which plays a major role in the Bayesian framework. It is a combination of the delayed rejection (DR) algorithm, which postpones the rejection and provides more sample candidates at each update step, and the adaptive Metropolis (AM) algorithm, which periodically adjusts the covariance of the proposal distribution based on the chain so far. For DR in all our experiments, we proposed a maximum of 2 candidates at each update step. The \textbf{FME} package in \textbf{R} was used for implementation. In the process, the `lsoda' function using C/C++ was exploited to compute the ODE solutions.

\item  \textbf{Hamiltonian Monte Carlo (HMC) }\\
Hamiltonian Monte Carlo originated from \citeasnoun**{duane1987hybrid} is an MCMC method using Hamiltonian dynamics. The HMC method introduces auxiliary momentum variables to transform the sampling problem into a fictitious Hamiltonian simulation with potential and kinetic energy. This strategy is known to be significantly better than Metropolis updates with a  random-walk proposal in the exploration of parameter space. For details, see \citeasnoun{neal2011mcmc}. The \textbf{R} package \textbf{rstan} provides an implementation for the ODE model. The No-U-Turn Sampler (NUTS) by \citeasnoun{hoffman2014no}, the default option of \textbf{rstan}, is a variant algorithm to choose the tuning parameters for HMC automatically.  We used it for simulation experiments. The package uses a C++ compiler.

\item  \textbf{Relaxed DEM with extended Liu and West filter (RDEM) }\\
 The idea of relaxing an ODE model to a state-space model was devised by \citeasnoun{lee2018inference}. 
They used a sequential Monte Carlo, the extended Liu and West filter of \citeasnoun{rios2013extended} for posterior computation. It was reported that RDEM is faster than PC and DRAM empirically; see \citeasnoun**{lee2018inference}. Since it has almost the same approximation process to the state-space model, we set its step size ($m$) and tuning parameter ($\tau$) the same as ours. The priors for the RDEM algorithm were
\begin{align*}
\bfx_0|\lambda&\sim\text{ N}_p(\bfy_0,\lambda^{-1}\bfI_p), \\
\lambda&\sim\text{ Gamma}(1,\ 1)
\end{align*} 
in all the experiments here, following the original article. These conjugate priors facilitate the sampling algorithm and differ from ours, which aims to make the cost function calculation easier.
The method was implemented using C++ code via the \textbf{Rcpp} package.

\ei

Our method is denoted by SSVB (state-space model with variational Bayes) and implemented using C++ code via the \textbf{Rcpp} package. The tuning parameters for SSVB, the step size $m$ and $\tau$, were determined by the procedure suggested in Appendix \ref{tuning}. This procedure is not absolute, but it can be a suitable recommendation method.

In each ODE model, 100 data sets were generated from one true model, and the estimates based on the five methods were compared. For the Bayesian methods, the mean of the posterior was taken as a point estimator.

For a fair comparison, all methods begin their algorithms at the same starting point. In other words, when the five methods need any starting point in their optimization or MCMC chains, their starting points are different across the data sets but the same across the methods in each data set. The starting points of ODE parameters are drawn from uniform distributions under the assumption that little prior information is available except for their roughly possible ranges. The starting points of $\bfx_0$ are selected using the (simple) cubic B-spline regression.

\subsection{FitzHugh-Nagumo model} 
\begin{figure}[b!]
	\centering
	\includegraphics[width=0.7\linewidth]{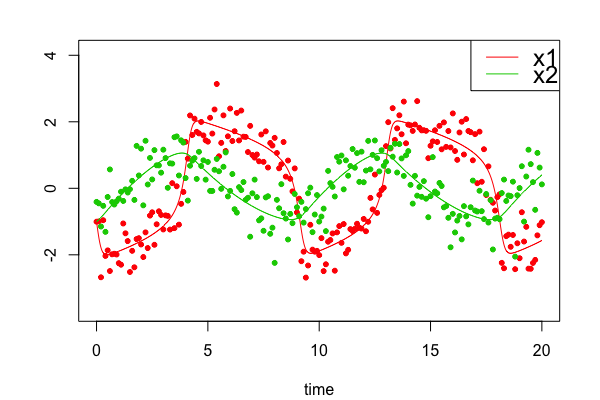}
	\caption{FitzHugh-Nagumo model of $\btheta = (0.2,\ 0.2,\ 3)^T$ and $\bfx_0 = (-1,\ -1)^T$. One of the data sets is plotted.}
	\label{fig:NagumoData}
\end{figure}
FitzHugh-Nagumo system of \citeasnoun{fitzhugh1961impulses} and \citeasnoun**{nagumo1965active} describes the interaction between the membrane voltage $x_1$ and the outwards current $x_2$ of a giant squid axon:
\begin{align*}
\dot{x}_1(t)=&\ \ \theta_3\left(x_1(t)-\frac{1}{3}x^3_1(t)+x_2(t)\right),\\
\dot{x}_2(t)=&-\frac{1}{\theta_3}\bigg(x_1(t)-\theta_1+\theta_2x_2(t)\bigg).
\end{align*}  
This system with two variables and three parameters was chosen as a moderately small ODE system since it has been regarded as a benchmark problem in numerous papers for ODE parameter estimation. From the true model of  $\btheta = (0.2,\ 0.2,\ 3)^T$ and $\bfx_0 = (-1,\ -1)^T$, 100  data sets  were generated with the error variance $1/\lambda=0.25$ along the 201 equidistant time points $ t_0=0,\ t_1=0.1,\,\ \cdots,\ t_{200}=20 $. The solution lines and one  data set  are shown in Figure~\ref{fig:NagumoData}.

\begin{figure}[b!]
	\centering
	\begin{subfigure}{0.48\linewidth}
		\includegraphics[width=\linewidth]{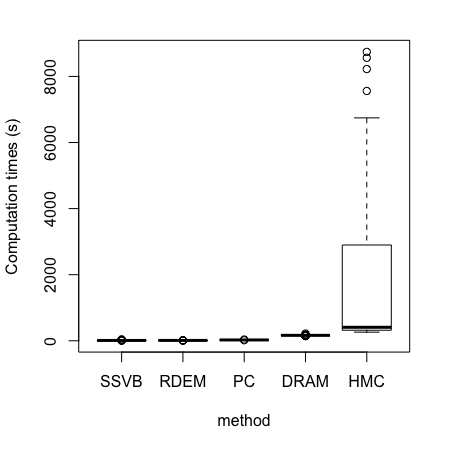}
	\end{subfigure}
	\begin{subfigure}{0.48\linewidth}
		\includegraphics[width=\linewidth]{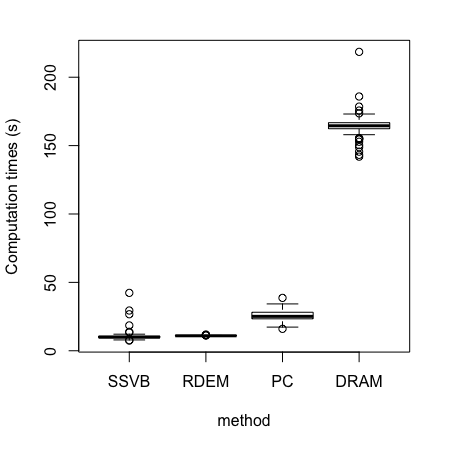}
	\end{subfigure}
	\caption{The boxplots of the running time of the inference for the 100 data sets from the FitzHugh-Nagumo model.  A comparison of all five methods on the left and four methods on the right excluding the slowest HMC algorithm. }
	\label{fig:NagumoTimes}
\end{figure} 
For the SSVB, the priors are :
\begin{align*}
\lambda&\sim \ \text{Gamma}(1,\ 1) \\
\btheta&\sim \text{Unif }\{(-0.8,0.8)\times(-0.8,0.8)\times(0,8)\}.
\end{align*}
The bounds for $\bfx_0$'s uniform priors are appropriately selected to be centered at the starting point of $\bfm_0$ (variational parameter for $\bfx_0$) from the cubic B-spline regression. With the tuning parameters $m=1$, $\tau=0.1^5$ from the procedure in Appendix \ref{tuning}, the variational parameters were optimized.

For the other methods, the PC method's smoothing parameter $\lambda =1,000$ was chosen from the FPE of Ellner (2007)\nocite{ellner2007commentary}. The DRAM and HMC were implemented with a chain size of 10,000, respectively, and for the DRAM, the proposal covariance was updated every 100 iterations. The first 5,000 iterations of the chains were discarded as burn-in. The RDEM algorithm was implemented with 20,000 particles.

\begin{figure}[hp]
	\centering
	\begin{subfigure}{0.9\linewidth}
		\includegraphics[width=\linewidth]{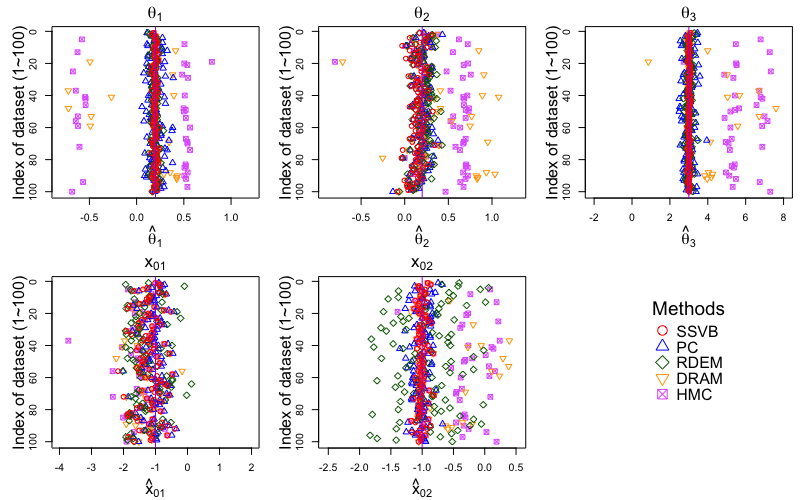}
		\caption{Comparison of the five methods.}
		\label{fig:NagumoEst5}
	\end{subfigure}
	\begin{subfigure}{0.9\linewidth}
		\includegraphics[width=\linewidth]{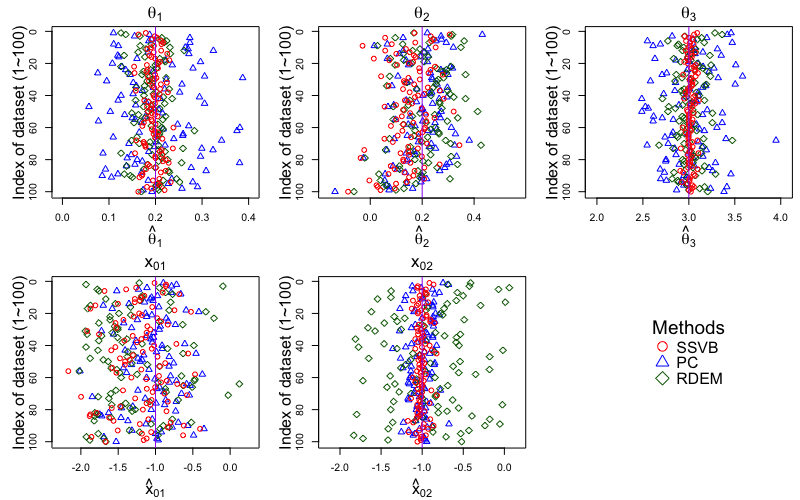}
		\caption{Comparison without the DRAM and HMC.}
		\label{fig:NagumoEst3}
	\end{subfigure}
	\caption{The resulting estimates from the five methods for the FitzHugh-Nagumo model. Each plot was centered on the true parameter value represented by a purple vertical line.}
	\label{fig:NagumoEst}
\end{figure}
Figure~\ref{fig:NagumoTimes} shows the boxplots of the running times of each method for 100 data sets. The SSVB and RDEM showed the fastest speed with an average time of 10.79 and 10.98 seconds, respectively, followed by the PC and DRAM in order with 25.75 seconds and 2.74 minutes. The HMC (NUTS) algorithm took the longest average time of 30.06 minutes. The exceptionally long average time of HMC is due to frequent convergence failures. 36 out of 100 estimates failed to converge, and the average time for these cases was 1.19 hours. The average time for the other 64 convergence cases was 6.68 minutes.

The estimates from the five methods on 100 data sets are plotted in Figure~\ref{fig:NagumoEst}. In the figure, the vertical axis represents the indices of the data sets, and the horizontal axis represents the resulting estimates for the data set. The true parameter value was represented by a purple vertical line in the center. The DRAM and HMC algorithms fairly often showed poor performance in inferring the ODE parameters, as shown in Figure~\ref{fig:NagumoEst5}. It appears that the DRAM suffers poor mixing, since the shape of the solution curve, and hence the likelihood, varies dramatically depending on the parameter combination. The HMC, known for better exploration in the parameter space, also frequently failed to reach the main convergence area depending on the starting point. Excluding the DRAM and HMC, Figure~\ref{fig:NagumoEst3} shows the others' results in more detail. We can visually confirm that the SSVB is overall the best method with the smallest variability as well as the  smallest  biases.

\begin{table}[b!]
	\begin{subtable}{\linewidth}
		\centering
		{\footnotesize \begin{tabular}{c|ccccc|c|c}
				\hline
				& SSVB & PC & RDEM & DRAM & HMC & min. & ratio \\ 
				\hline
				$\theta_1 $ & 0.0150 & 0.0630 & 0.0255 & 0.0733 & 0.1915 & SSVB & 1.7017 \\ 
				$\theta_2 $ & 0.0740 & 0.0731 & 0.0775 & 0.1562 & 0.2069 & PC & 0.9883 \\ 
				$\theta_3 $ & 0.0335 & 0.2256 & 0.1121 & 0.3386 & 1.0365 & SSVB & 3.3476 \\ 
				\hdashline
				$x_{01}$ & 0.3291 & 0.2730 & 0.4915 & 0.3245 & 0.3880 & PC & 0.8296 \\ 
				$x_{02}$ & 0.0522 & 0.1128 & 0.3929 & 0.1761 & 0.3285 & SSVB & 2.1603 \\ 
				\hline
		\end{tabular}}
		\caption{Mean absolute bias}
	\end{subtable}
	\hfill \\
	\hfill \\
	\hfill
	\begin{subtable}{\linewidth}
		\centering
		{\footnotesize \begin{tabular}{c|ccccc|c|c}
				\hline
				& SSVB & PC & RDEM & DRAM & HMC & min. & ratio\\ 
				\hline
				$\theta_1 $ & 0.0187 & 0.0767 & 0.0321 & 0.1910 & 0.3348 & SSVB & 1.7166 \\ 
				$\theta_2 $ & 0.0794 & 0.0920 & 0.0958 & 0.2667 & 0.2605 & SSVB & 1.1593 \\ 
				$\theta_3 $ & 0.0415 & 0.2818 & 0.1419 & 0.8813 & 1.4810 & SSVB & 3.4204 \\ 
				\hdashline
				$x_{01}$ & 0.3712 & 0.3482 & 0.4948 & 0.3658 & 0.4561 & PC & 0.9381 \\ 
				$x_{02}$ & 0.0684 & 0.1385 & 0.4713 & 0.3429 & 0.4305 & SSVB & 2.0230 \\  
				\hline
		\end{tabular}}
		\caption{Sample standard deviation}
	\end{subtable}
	\caption{MAB and SSD for the 100 estimates in the FitzHugh-Nagumo model.}
	\label{tab:NagumoBiasStd}
\end{table}

More specifically, Table~\ref{tab:NagumoBiasStd} shows the mean absolute bias (MAB) and the sample standard deviation (SSD) for the 100 estimates. The last column, `ratio', denotes the ratio of the SSVB to the best of the other methods. For example, for the MAB of $\theta_1$, the best of the others is the RDEM's 0.0255, and it is 1.7017 times the SSVB' 0.0150. Overall, we can see that the SSVB provides a fairly good estimate compared to the others. It showed the smallest SSD for all parameters except for $x_{01}$.

\begin{figure}[t!]
	\centering
	\includegraphics[width=0.9\linewidth]{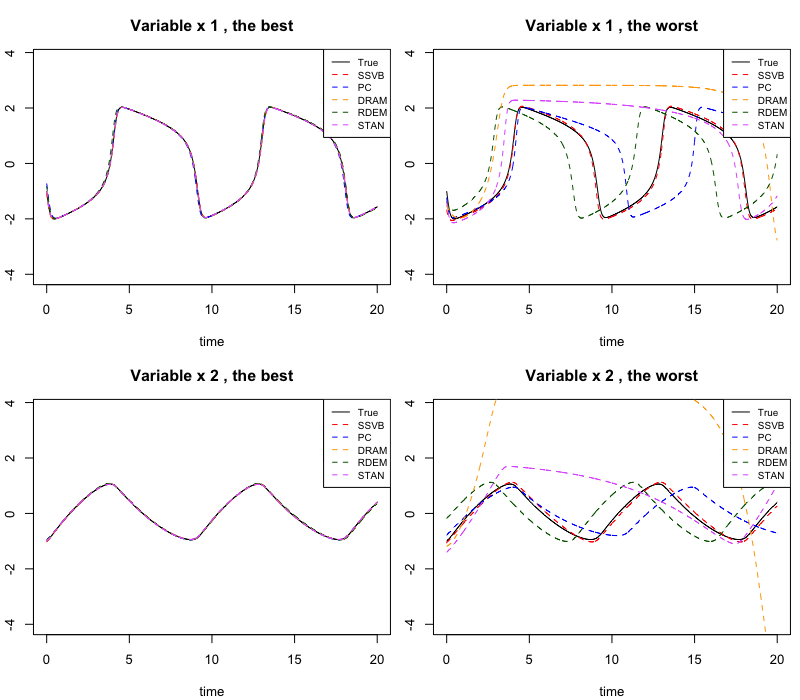}
	\caption{The estimated solution curves with their best/worst estimates ($\hat{\btheta}$, $\hat{\bfx}_0$) from the 100 data sets, of the FitzHugh-Nagumo model, for each method.}
	\label{fig:NagumoBestWorst}
\end{figure}

Finally, we checked the estimated solution curves with their best/worst estimates ($\hat{\btheta}$, $\hat{\bfx}_0$) from the 100 data sets.
In Figure~\ref{fig:NagumoBestWorst}, all the methods provided their best curves that closely match the solution curve with the true parameter values. However, in the worst estimate, only the SSVB method produced curves that are similar to the true solution. This confirms that the SSVB method provides a relatively very stable estimator. Note that the best/worst estimates were selected based on the deviation (sum of squares) from the true solution curve at the observation times.

\subsection{Lorenz-96 model}
As a big ODE model for comparing the performance,  we adopted the Lorenz-96 model by  \citeasnoun{lorenz1995predictability}. The Lorenz-96 model is a toy model for an unspecified meteorological quantity such as temperature or concentration of a substance. The reason for choosing the Lorenze-96 model as a testing model is that one can enlarge the model as desired by increasing the number of variables $p\ (\geq3)$. 

\begin{figure}[b]
	\centering
	\includegraphics[width=0.7\linewidth]{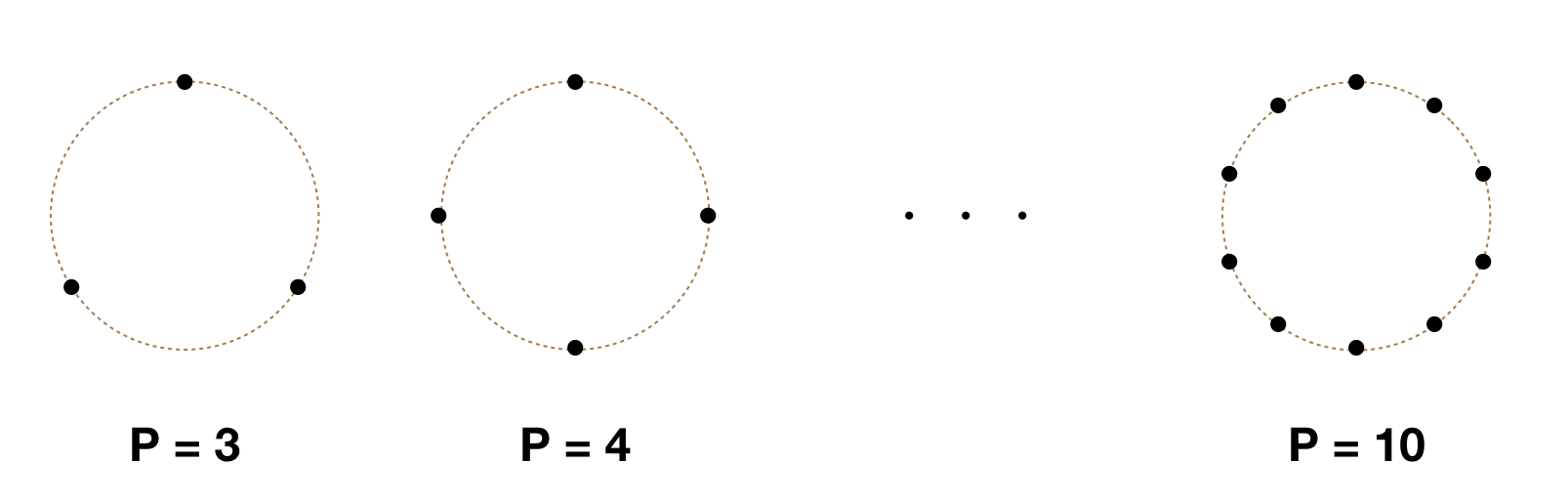}
	\caption{The scalability of the Lorenz-96 model.}
	\label{fig:Lorenz96circle}
\end{figure}

The $p$ scalar variables at equally spaced sites around a latitude circle, as shown in Figure \ref{fig:Lorenz96circle}, have the relationship given by $$  \frac{dX_j}{dt} = (X_{j+1}-X_{j-2})X_{j-1}-X_j+F,\qquad\text{for }\ j=1,\  \dots,\ p. $$
According to the circular structure, $X_{-1}=X_{p-1}$, $X_0=X_p$, and $X_{p+1}=X_1$. The quadratic terms and the linear terms correspond to advection and dissipation, respectively. The only ODE parameter is the constant term $F$, corresponding to external forcing.

In fact, the above model is a reduced model letting the coefficients of the quadratic and linear terms be 1. In this simulation study,  for the purpose of  performance   comparison in a large model with many parameters and variables, we use the following model: 
$$  \frac{dX_j}{dt} = \theta_{1j}(X_{j+1}-X_{j-2})X_{j-1}-\theta_{2j}X_j+\theta_{3j},\qquad\text{for }\ j=1,\  \dots,\ p.$$
If the number of variables is $p$,  the number of ODE parameters is $q=3p$; thus, the total number of parameters to be estimated is $q+p=4p$, including the initial states. If the model has 10 variables, for example, a total of 40 parameters need to be inferred. 
 
In this paper, we conducted experiments for the model with $p=4$ and $10$, respectively. The true parameter values are $(\theta_{1j}, \theta_{2j}, \theta_{3j})=(1, 1, 8)$ for all $j=1,\  \dots,\ p$ in both cases, reflecting the original model ($\theta_{1j}=\theta_{2j}=1$) and  guaranteeing the chaotic behaviour ($\theta_{3j}=8$); see \citeasnoun{lorenz1998optimal}. 

The DRAM and HMC algorithms were excluded from these models due to their extremely slow computation and poor performance. For the $p=4$ model, we tried the DRAM with a chain size of 100,000. Of the 31 cases, 14 were aborted since they failed to compute numerical ODE solutions during execution. The remaining 17 cases took an average of 60.72 minutes, and only one case converged. The HMC was tried with a chain size of 50,000. The calculation took an average of 29.96 hours, and convergence occurred in 5 out of 6 cases. Also, 2 out of 5 converged cases took more than 6 hours to reach the main convergence area. Considering that the other three methods produced fairly good estimate results within an average of less than 1 minute, these two MCMC methods were judged to be inefficient and excluded.

\begin{figure}[b!]
	\centering
	\includegraphics[width=0.7\linewidth]{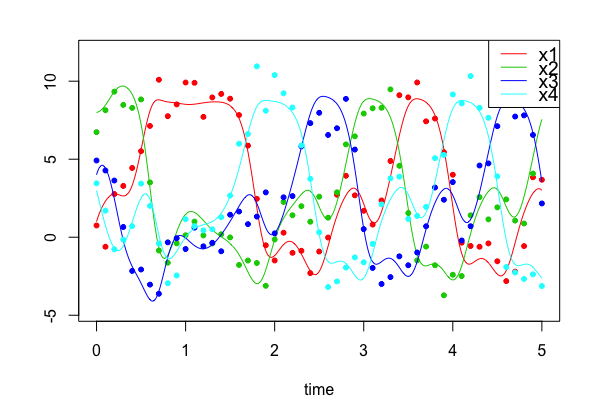}
	\caption{Lorenz-96 model with 4 variables of $\bfx_0 = (1,\ 8,\ 4,\ 3)^T$. One of the data sets is plotted.}
	\label{fig:4LorenzData}
\end{figure}

\subsubsection{Lorenz-96 model with 4 variables}
First, we generated 100 data sets from the Lorenz-96 model with 4 variables. The error variance was $1/\lambda=1$ and the initial states  $\bfx_0 = (1,\ 8,\ 4,\ 3)^T$ were randomly selected. Observations are made at the 51 equidistant time points $t_0=0,\ t_1=0.1,\ \cdots,\ t_{50}=5$. The solution lines and one data set are  shown in Figure~\ref{fig:4LorenzData}. The total number of parameters to be estimated is $p+q=4+12=16$.

The priors for the SSVB were :
\begin{align*}
\lambda\sim& \ \text{Gamma}(1,\ 1), \\
\btheta_j\sim&\ \text{Unif }\{(0,2)\times(0,2)\times(0,16)\},\ \text{ for } j=1,\dots,4,
\end{align*}
and the uniform priors for $\bfx_0$ were properly selected from the data set. The procedure in Appendix \ref{tuning} recommended $m=2$ and $\tau=0.1^4$. The smoothing parameter of the PC method was set by $\lambda =10,000$ based on the FPE of \citeasnoun{ellner2007commentary}. The RDEM algorithm was implemented with 50,000 particles. 

\begin{figure}[b!]
	\centering
	\includegraphics[width=0.5\linewidth]{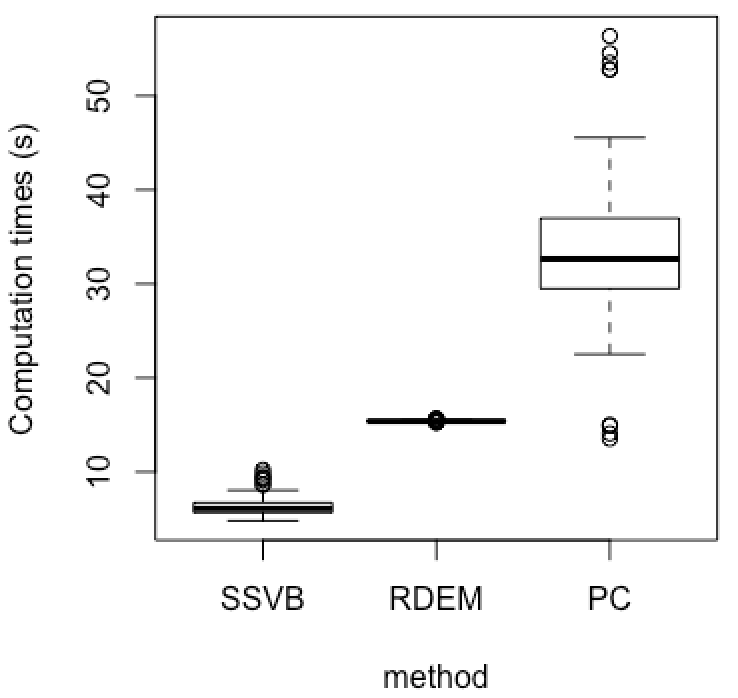}
	\caption{The boxplots of the running time of the inference for the 100 data sets from the Lorenz-96 model with 4 variables.}
	\label{timeLorenz4}
\end{figure}
\begin{figure}[b!]
	\includegraphics[width=\linewidth]{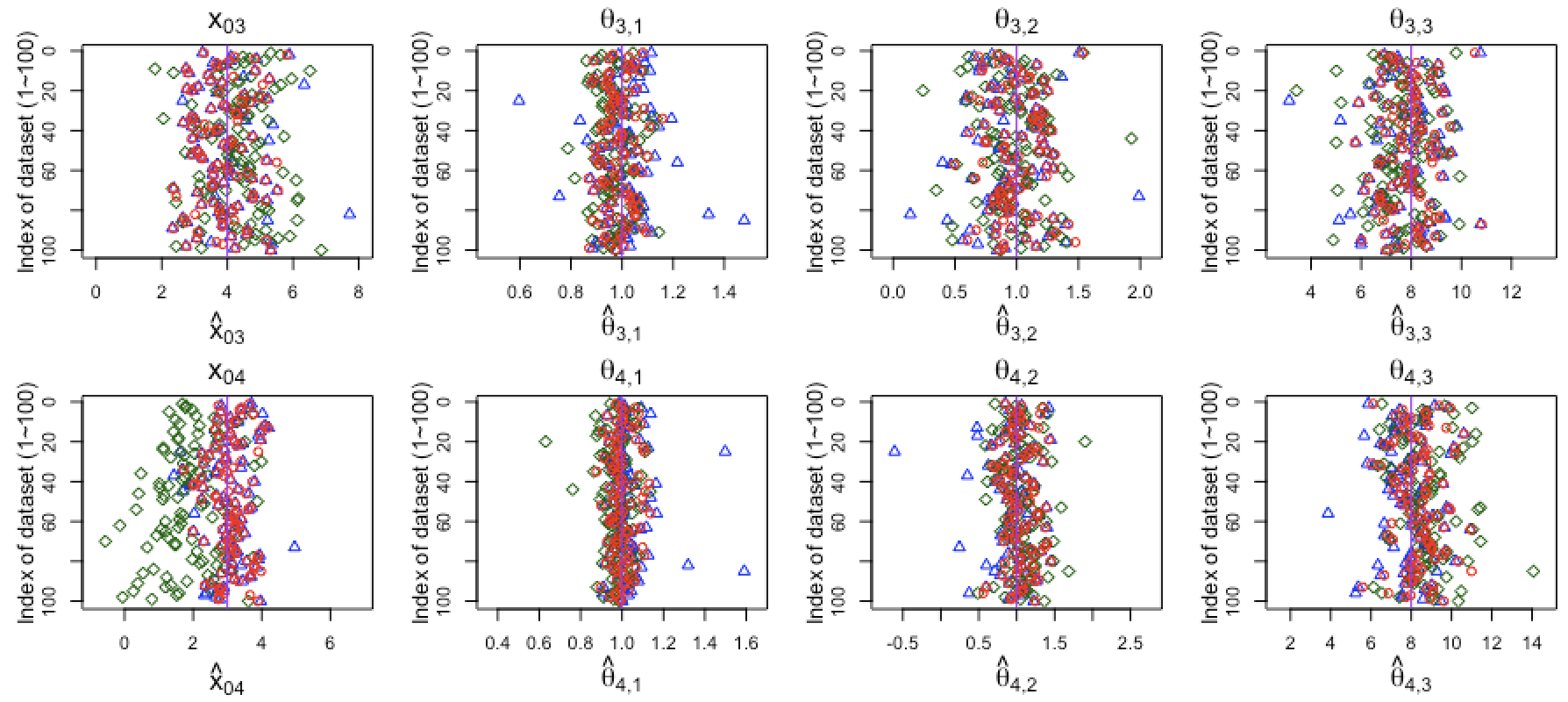}
	\caption{The resulting estimates from the three methods for the Lorenz-96 model with 4 variables. \textit{Red circle}: SSVB, \textit{Blue triangle}: PC, \textit{Green diamond}: RDEM. Each plot was centered on a true parameter value represented by a purple vertical line. Results were shown for only 8 parameters, but the other 8 parameters also showed similar patterns.}
	\label{fig:4LorenzEst}
\end{figure}

Again, the computation speed of the SSVB was the fastest, as shown in Figure~\ref{timeLorenz4}. The mean time of 6.38 seconds of the SSVB was followed by 15.40 seconds of the RDEM and 33.36 seconds of the PC.

\begin{table}[b!]
	\begin{subtable}{0.47\textwidth}
		\centering
		{\footnotesize \begin{tabular}{c|ccc|c|c}
				\hline
				& SSVB & PC & RDEM & min. & ratio. \\ 
				\hline
				$\theta_{1,1}$ & 0.0460 & 0.0799 & 0.0569 & SSVB & 1.2371 \\ 
				$\theta_{1,2}$ & 0.1172 & 0.1414 & 0.1425 & SSVB & 1.2059 \\ 
				$\theta_{1,3}$ & 0.9275 & 1.0487 & 1.0879 & SSVB & 1.1307 \\ 
				$\theta_{2,1}$ & 0.0382 & 0.0581 & 0.0478 & SSVB & 1.2506 \\ 
				$\theta_{2,2}$ & 0.1659 & 0.1809 & 0.2392 & SSVB & 1.0906 \\ 
				$\theta_{2,3}$ & 0.7625 & 0.9102 & 1.1903 & SSVB & 1.1937 \\ 
				$\theta_{3,1}$ & 0.0511 & 0.0734 & 0.0598 & SSVB & 1.1716 \\ 
				$\theta_{3,2}$ & 0.1699 & 0.2078 & 0.2013 & SSVB & 1.1852 \\ 
				$\theta_{3,3}$ & 0.7544 & 0.9515 & 0.9462 & SSVB & 1.2542 \\ 
				$\theta_{4,1}$ & 0.0423 & 0.0567 & 0.0533 & SSVB & 1.2598 \\ 
				$\theta_{4,2}$ & 0.1542 & 0.1930 & 0.1871 & SSVB & 1.2132 \\ 
				$\theta_{4,3}$ & 0.8622 & 0.9440 & 1.1372 & SSVB & 1.0949 \\  
				\hdashline
				$x_{01}$ & 0.4280 & 0.4712 & 1.3701 & SSVB & 1.1011 \\ 
				$x_{02}$ & 0.3873 & 0.5338 & 0.7132 & SSVB & 1.3782 \\ 
				$x_{03}$ & 0.6681 & 0.7406 & 0.8444 & SSVB & 1.1085 \\ 
				$x_{04}$ & 0.4028 & 0.4852 & 1.2484 & SSVB & 1.2045 \\
				\hline
		\end{tabular}}
		\caption{Mean absolute bias}
	\end{subtable}
	\hfill
	\begin{subtable}{0.47\textwidth}
		{\footnotesize \begin{tabular}{c|ccc|c|c}
				\hline
				& SSVB & PC & RDEM & min. & ratio. \\ 
				\hline
				$\theta_{1,1}$ & 0.0554 & 0.1589 & 0.0692 & SSVB & 1.2480 \\ 
				$\theta_{1,2}$ & 0.1468 & 0.1980 & 0.1926 & SSVB & 1.3122 \\ 
				$\theta_{1,3}$ & 1.0833 & 1.4475 & 1.3647 & SSVB & 1.2597 \\ 
				$\theta_{2,1}$ & 0.0495 & 0.1161 & 0.0608 & SSVB & 1.2286 \\ 
				$\theta_{2,2}$ & 0.1836 & 0.2379 & 0.2664 & SSVB & 1.2957 \\ 
				$\theta_{2,3}$ & 0.9328 & 1.2679 & 1.3583 & SSVB & 1.3593 \\ 
				$\theta_{3,1}$ & 0.0604 & 0.1054 & 0.0684 & SSVB & 1.1324 \\ 
				$\theta_{3,2}$ & 0.2060 & 0.2639 & 0.2613 & SSVB & 1.2689 \\ 
				$\theta_{3,3}$ & 0.9532 & 1.2086 & 1.1554 & SSVB & 1.2121 \\ 
				$\theta_{4,1}$ & 0.0552 & 0.0993 & 0.0708 & SSVB & 1.2830 \\ 
				$\theta_{4,2}$ & 0.1909 & 0.2861 & 0.2422 & SSVB & 1.2689 \\ 
				$\theta_{4,3}$ & 1.0505 & 1.2149 & 1.3535 & SSVB & 1.1566 \\ 
				\hdashline
				$x_{01}$ & 0.4881 & 0.6063 & 0.9403 & SSVB & 1.2420 \\ 
				$x_{02}$ & 0.4801 & 0.9362 & 0.8861 & SSVB & 1.8455 \\ 
				$x_{03}$ & 0.8047 & 0.9389 & 1.0082 & SSVB & 1.1668 \\ 
				$x_{04}$ & 0.5074 & 0.6228 & 0.9270 & SSVB & 1.2274 \\ 
				\hline
		\end{tabular}}
		\caption{Sample standard deviation}
	\end{subtable}
	\caption{MAB and SSD for the 100 estimates in the Lorenz-96 model with 4 variables.}
	\label{tab:4LorenzBiasStd}
\end{table} 

The inference results are plotted in Figure~\ref{fig:4LorenzEst}. Of the 16 parameters, only 8 results associated with $X_3$ and $X_4$ were displayed, but the other 8 parameters also showed similar patterns. Of the graphs arranged in four columns, the first column represents the results of the initial states $\bfx_0$, and the other three columns represent the results for $\btheta_j=(1,1,8)$ for $j=1, \dots, 4$. The RDEM algorithm tended to be biased in estimating the $\bfx_0$. Some of the 100 data sets led the PC method to make poor estimates. This shows that the PC method can fail depending on the given data set and the starting point of the algorithm. In the next experiment with 10 variables, we can confirm that this tendency increases together with the number of estimated parameters.

The results of the MAB and SSD are shown in Table~\ref{tab:4LorenzBiasStd}. In both MAB and SSD, the SSVB method had the minimum values for all parameters. For the SSD, the ratio with the second place reached a maximum of 1.85. The results showed the relative stability of the SSVB method.

\begin{figure}[t!]
	\centering
	\includegraphics[width=0.7\linewidth]{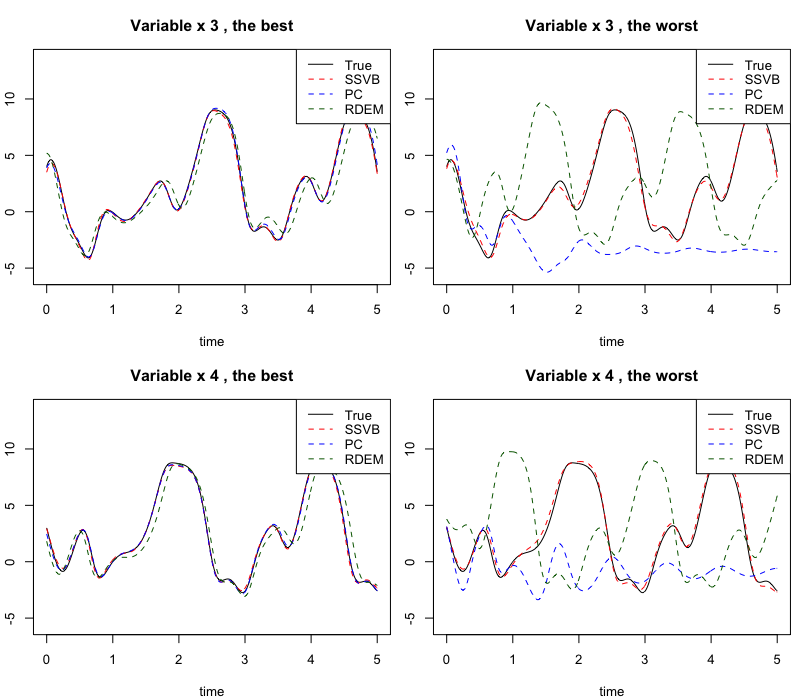}
	\caption{The estimated solution curves for $X_3$ and $X_4$ with their best/worst estimates ($\hat{\btheta}$, $\hat{\bfx}_0$) from the 100 data sets, of the Lorenz-96 model with 4 variables, for each method. $X_1$ and $X_2$ also showed similar patterns. }
	\label{fig:4LorenzBestWorst}
\end{figure} 
The best/worst estimated curves for $X_3$ and $X_4$ in Figure~\ref{fig:4LorenzBestWorst} also support the stability of the SSVB. All the methods yielded estimated curves very close to the true curves in their best estimate, but the worst estimated curves were rather different. The SSVB method was the only one that follows the true solution curve well even in the worst case. Although omitted, the results for $X_1$ and $X_2$ also had the same patterns. The stability of the SSVB can be confirmed more clearly in the following big model, the Lorenz-96 model with 10 variables.

\newpage
\subsubsection{Lorenz-96 model with 10 variables}
\begin{figure}[t!]
	\centering
	\includegraphics[width=0.8\linewidth]{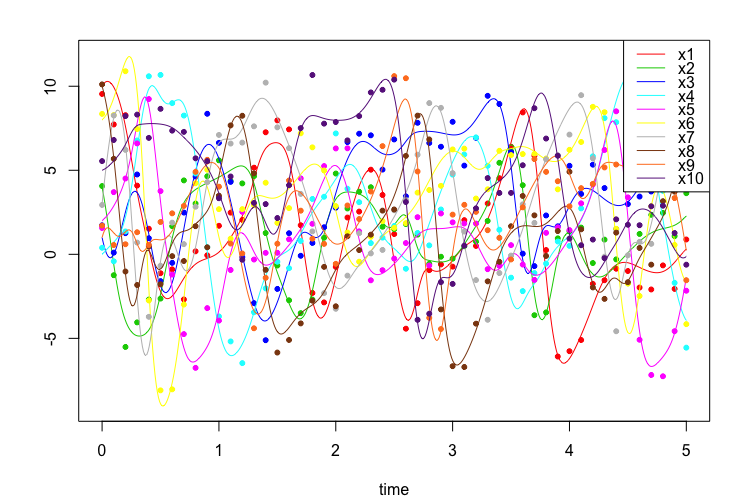}
	\caption{Lorenz-96 model with 10 variables of $\bfx_0 = (10,\ 4,\ 1,\ 0,\ 2,\ 8,\ 3,\ 10,\ 1,\ 5)^T$. One of the data sets is plotted.}
	\label{fig:10LorenzData}
\end{figure}

As the number of parameters to be estimated reaches $p+q=10+30=40$, the inference accuracy was expected to be worse than the model with 4 variables.  Along the 51 time points $t_0=0,\ t_1=0.1,\ \cdots,\ t_{50}=5$, a total of 100 data sets were generated from the true solution curve with $\bfx_0 = (10,\ 4,\ 1,\ 0,\ 2,\ 8,\ 3,\ 10,\ 1,\ 5)^T$. As in the previous simulation, the observation errors with variance $1/\lambda =1$ are added. The true solution curves and one data set are shown in Figure~\ref{fig:10LorenzData}.

All the priors and settings for the SSVB method were the same as in the previous case of 4 variables except for the step size $m=3$. The smoothing parameter of the PC method based on the FPE of \citeasnoun{ellner2007commentary} also has the same value $\lambda=10,000$ as before. The size of particles of the RDEM algorithm was increased to 200,000 so as to afford the big model.

\begin{figure}[b!]
	\centering
	\includegraphics[width=0.5\linewidth]{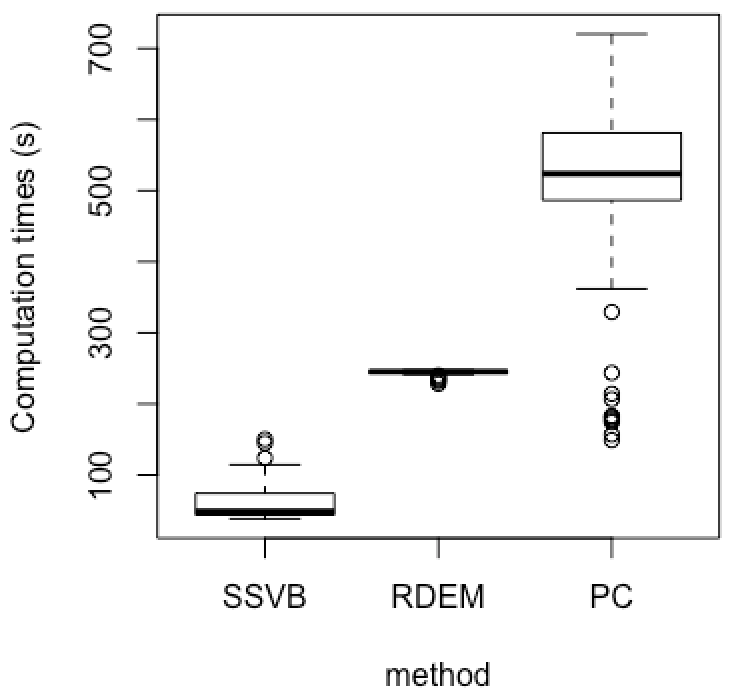}
	\caption{The boxplots of the running time of the inference for the 100 data sets from the Lorenz-96 model with 10 variables.}
	\label{fig:10LorenzTimes}
\end{figure}

\begin{figure}[b!]
	\includegraphics[width=\linewidth]{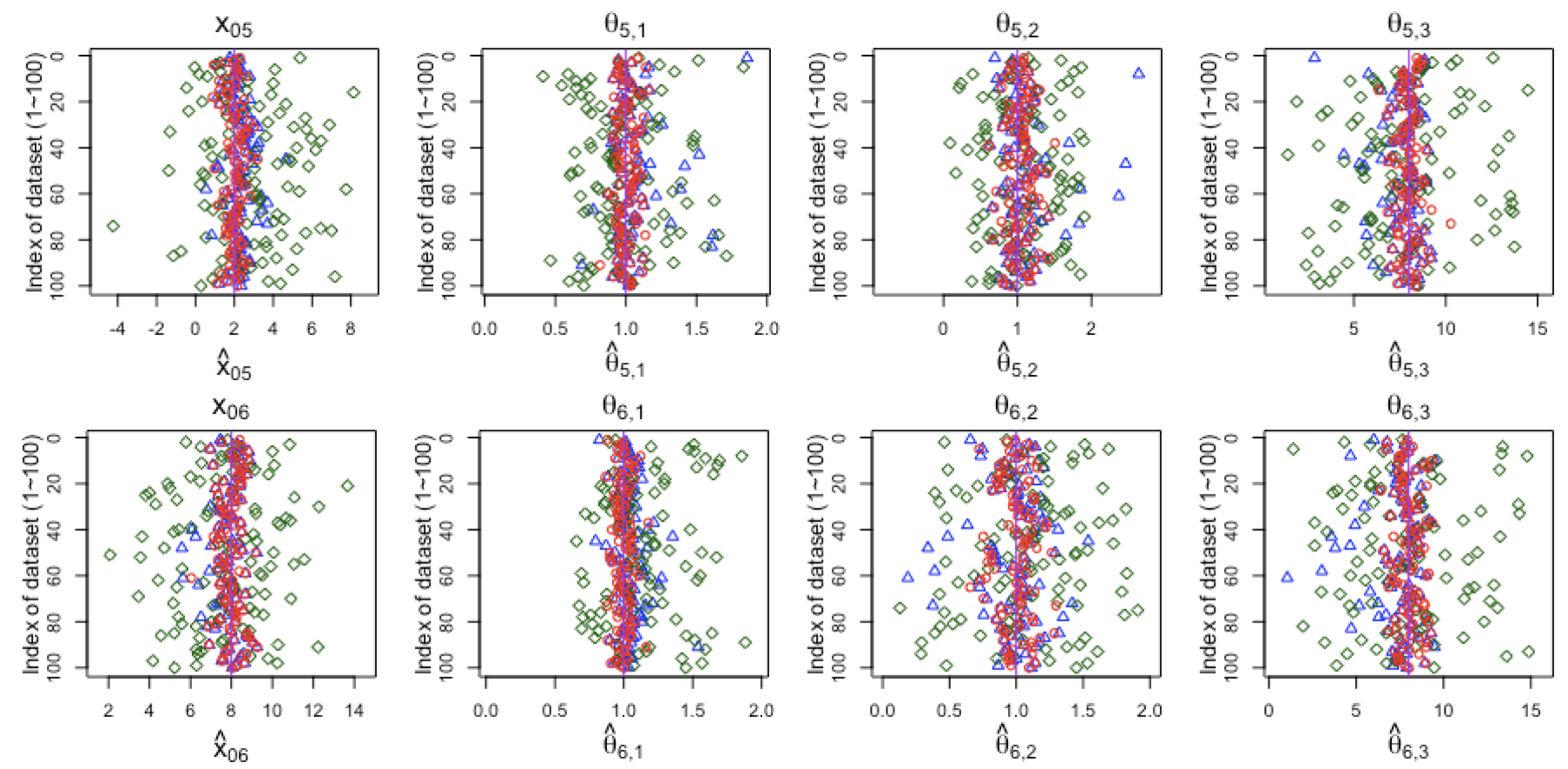}
	\caption{The resulting estimates related to $X_5$ and $X_{6}$ from the three methods for the Lorenz-96 model with 10 variables. \textit{Red circle}: SSVB, \textit{Blue triangle}: PC, \textit{Green diamond}: RDEM. Each plot was centered on a true parameter value represented by a purple vertical line. Results were shown for only 8 parameters, but the other 32 parameters also showed similar patterns.}
	\label{fig:10LorenzEst}
\end{figure}

The running times of the methods are plotted in Figure~\ref{fig:10LorenzTimes}. The mean time of the SSVB was the fastest at 54.33 seconds, followed in turn by the RDEM of 3.66 minutes and the PC of 5.98 minutes. As a result, the SSVB showed the fastest inference speed in all the simulation experiments.

What we should notice in the big model experiment is, in fact, the accuracy of the estimation rather than the speed of computation. Figure~\ref{fig:10LorenzEst} shows the resulting estimates from the three methods. The number of severe failures was greater than before in both PC and RDEM algorithms. On the other hand, the SSVB method tended to be relatively close to the true parameter values, not significantly affected by the data set. These overall trends were the same for the 32 omitted parameters.

\begin{table}[t]
	\begin{subtable}{0.47\textwidth} 
		\centering
		{\footnotesize \begin{tabular}{c|ccc|c|c}
				\hline
				& SSVB & PC & RDEM & min. & ratio. \\ 
				\hline
				$\theta_{1,1}$ & 0.0361 & 0.0619 & 0.2182 & SSVB & 1.7148 \\ 
  				$\theta_{1,2}$ & 0.1388 & 0.2074 & 0.4215 & SSVB & 1.4938 \\ 
  				$\theta_{1,3}$ & 0.4220 & 0.6689 & 2.8240 & SSVB & 1.5849 \\ 
  				$\theta_{2,1}$ & 0.0413 & 0.1177 & 0.1937 & SSVB & 2.8481 \\ 
  				$\theta_{2,2}$ & 0.1644 & 0.2527 & 0.3956 & SSVB & 1.5373 \\ 
  				$\theta_{2,3}$ & 0.5753 & 0.8469 & 2.6669 & SSVB & 1.4723 \\ 
  				$\theta_{3,1}$ & 0.0530 & 0.0937 & 0.2914 & SSVB & 1.7670 \\ 
  				$\theta_{3,2}$ & 0.1402 & 0.1505 & 0.4222 & SSVB & 1.0739 \\ 
  				$\theta_{3,3}$ & 0.8667 & 0.9371 & 2.6462 & SSVB & 1.0812 \\ 
  				$\theta_{4,1}$ & 0.0403 & 0.0539 & 0.2583 & SSVB & 1.3361 \\ 
  				$\theta_{4,2}$ & 0.0954 & 0.1237 & 0.3914 & SSVB & 1.2967 \\ 
  				$\theta_{4,3}$ & 0.3466 & 0.4795 & 2.5656 & SSVB & 1.3833 \\ 
  				$\theta_{5,1}$ & 0.0553 & 0.1002 & 0.2398 & SSVB & 1.8105 \\ 
  				$\theta_{5,2}$ & 0.1444 & 0.2088 & 0.4074 & SSVB & 1.4463 \\ 
  				$\theta_{5,3}$ & 0.4527 & 0.7136 & 2.4929 & SSVB & 1.5763 \\ 
  				$\theta_{6,1}$ & 0.0474 & 0.0735 & 0.2483 & SSVB & 1.5511 \\ 
  				$\theta_{6,2}$ & 0.1003 & 0.1538 & 0.3437 & SSVB & 1.5337 \\ 
  				$\theta_{6,3}$ & 0.5557 & 0.9886 & 2.5301 & SSVB & 1.7791 \\ 
  				$\theta_{7,1}$ & 0.0395 & 0.0549 & 0.2381 & SSVB & 1.3909 \\ 
  				$\theta_{7,2}$ & 0.1166 & 0.1511 & 0.4100 & SSVB & 1.2958 \\  
		\end{tabular}}
	\end{subtable}
	\hfill
	\begin{subtable}{0.47\textwidth}
		{\footnotesize \begin{tabular}{c|ccc|c|c}
				& SSVB & PC & RDEM & min. & ratio. \\ 
				\hline
  				$\theta_{7,3}$ & 0.4689 & 0.9767 & 3.0383 & SSVB & 2.0829 \\ 
  				$\theta_{8,1}$ & 0.0418 & 0.0818 & 0.2128 & SSVB & 1.9562 \\ 
  				$\theta_{8,2}$ & 0.1646 & 0.1890 & 0.3757 & SSVB & 1.1485 \\ 
  				$\theta_{8,3}$ & 0.3417 & 0.5310 & 2.9875 & SSVB & 1.5540 \\ 
  				$\theta_{9,1}$ & 0.0462 & 0.0759 & 0.3280 & SSVB & 1.6431 \\ 
  				$\theta_{9,2}$ & 0.2121 & 0.2483 & 0.4184 & SSVB & 1.1704 \\ 
  				$\theta_{9,3}$ & 0.9793 & 1.2501 & 2.7519 & SSVB & 1.2765 \\ 
  				$\theta_{10,1}$ & 0.0498 & 0.0758 & 0.2698 & SSVB & 1.5212 \\ 
  				$\theta_{10,2}$ & 0.1099 & 0.1725 & 0.4179 & SSVB & 1.5692 \\ 
  				$\theta_{10,3}$ & 0.7858 & 0.8878 & 3.1258 & SSVB & 1.1298 \\ 
				\hdashline
  				$x_{01}$ & 0.3755 & 0.4585 & 1.9858 & SSVB & 1.2211 \\ 
  				$x_{02}$ & 0.4065 & 0.4779 & 3.1649 & SSVB & 1.1758 \\ 
  				$x_{03}$ & 0.5104 & 0.6412 & 1.8294 & SSVB & 1.2562 \\ 
  				$x_{04}$ & 0.2480 & 0.3235 & 1.7746 & SSVB & 1.3044 \\ 
  				$x_{05}$ & 0.3415 & 0.5114 & 1.9504 & SSVB & 1.4977 \\ 
  				$x_{06}$ & 0.4469 & 0.5581 & 1.8704 & SSVB & 1.2489 \\ 
  				$x_{07}$ & 0.5766 & 0.6889 & 2.3787 & SSVB & 1.1948 \\ 
  				$x_{08}$ & 0.4319 & 0.5577 & 2.6107 & SSVB & 1.2912 \\ 
  				$x_{09}$ & 0.5674 & 0.5565 & 2.0066 & PC & 0.9809 \\ 
  				$x_{010}$ & 0.3330 & 0.4869 & 1.9900 & SSVB & 1.4619 \\ 
				\hline
		\end{tabular}}
	\end{subtable}
	\caption{Mean absolute bias for the 100 estimates in the Lorenz-96 model with 10 variables.}
	\label{tab:10LorenzBias}
\end{table}
Actually, the MABs of the SSVB were the smallest for all the parameters except $x_{09}$ as shown in Table~\ref{tab:10LorenzBias}. The ratio with the second place reached 2.8481 at the maximum, and the others also showed substantial differences. For the SSD in which the SSVB showed the smallest values for all the parameters, the differences became more severe. Table~\ref{tab:10LorenzStd} shows that the maximum ratio reached 4.5534, and several other ratios exceeded 2. 

\begin{table}[t]
	\begin{subtable}{0.47\textwidth}
		\centering
		{\footnotesize \begin{tabular}{c|ccc|c|c}
				\hline
				& SSVB & PC & RDEM & min. & ratio. \\ 
				\hline
  				$\theta_{1,1}$ & 0.0441 & 0.0988 & 0.2727 & SSVB & 2.2382 \\ 
    			$\theta_{1,2}$ & 0.1654 & 0.2755 & 0.4867 & SSVB & 1.6653 \\ 
    			$\theta_{1,3}$ & 0.4800 & 0.9040 & 3.4055 & SSVB & 1.8833 \\ 
    			$\theta_{2,1}$ & 0.0496 & 0.2897 & 0.2261 & SSVB & 4.5534 \\ 
    			$\theta_{2,2}$ & 0.2039 & 0.4800 & 0.4666 & SSVB & 2.2884 \\ 
    			$\theta_{2,3}$ & 0.7094 & 1.2944 & 3.2228 & SSVB & 1.8246 \\ 
   				$\theta_{3,1}$ & 0.0664 & 0.1299 & 0.3030 & SSVB & 1.9572 \\ 
   				$\theta_{3,2}$ & 0.1524 & 0.2047 & 0.4803 & SSVB & 1.3437 \\ 
   				$\theta_{3,3}$ & 1.0227 & 1.3611 & 3.2660 & SSVB & 1.3309 \\ 
    			$\theta_{4,1}$ & 0.0493 & 0.0727 & 0.3028 & SSVB & 1.4751 \\ 
    			$\theta_{4,2}$ & 0.1196 & 0.1733 & 0.4712 & SSVB & 1.4494 \\ 
    			$\theta_{4,3}$ & 0.4413 & 0.6431 & 3.2410 & SSVB & 1.4573 \\ 
   				$\theta_{5,1}$ & 0.0667 & 0.1647 & 0.2959 & SSVB & 2.4673 \\ 
   				$\theta_{5,2}$ & 0.1736 & 0.3318 & 0.4782 & SSVB & 1.9111 \\ 
   				$\theta_{5,3}$ & 0.5993 & 1.0250 & 3.0979 & SSVB & 1.7103 \\ 
   				$\theta_{6,1}$ & 0.0593 & 0.1010 & 0.2890 & SSVB & 1.7016 \\ 
    			$\theta_{6,2}$ & 0.1273 & 0.2168 & 0.4140 & SSVB & 1.7026 \\ 
    			$\theta_{6,3}$ & 0.6933 & 1.4235 & 3.1694 & SSVB & 2.0532 \\ 
    			$\theta_{7,1}$ & 0.0497 & 0.0795 & 0.1551 & SSVB & 1.5987 \\ 
   				$\theta_{7,2}$ & 0.1456 & 0.1885 & 0.4699 & SSVB & 1.2949 \\ 
		\end{tabular}}
	\end{subtable}
	\hfill
	\begin{subtable}{0.47\textwidth}
		{\footnotesize \begin{tabular}{c|ccc|c|c}
				& SSVB & PC & RDEM & min. & ratio. \\ 
				\hline
    			$\theta_{7,3}$ & 0.5868 & 1.3677 & 3.2604 & SSVB & 2.3308 \\ 
    			$\theta_{8,1}$ & 0.0535 & 0.1330 & 0.2689 & SSVB & 2.4861 \\ 
   				$\theta_{8,2}$ & 0.2065 & 0.2472 & 0.4285 & SSVB & 1.1971 \\ 
   				$\theta_{8,3}$ & 0.4376 & 0.7454 & 3.4990 & SSVB & 1.7034 \\ 
   				$\theta_{9,1}$ & 0.0571 & 0.1146 & 0.3010 & SSVB & 2.0072 \\ 
    			$\theta_{9,2}$ & 0.2651 & 0.3155 & 0.4677 & SSVB & 1.1903 \\ 
    			$\theta_{9,3}$ & 1.1698 & 1.4806 & 2.9127 & SSVB & 1.2657 \\ 
    			$\theta_{10,1}$ & 0.0620 & 0.1126 & 0.3362 & SSVB & 1.8170 \\ 
    			$\theta_{10,2}$ & 0.1364 & 0.2651 & 0.4915 & SSVB & 1.9437 \\ 
   				$\theta_{10,3}$ & 0.9393 & 1.1058 & 3.7625 & SSVB & 1.1772 \\  
				\hdashline
    			$x_{01}$ & 0.4742 & 0.6116 & 2.5345 & SSVB & 1.2897 \\ 
    			$x_{02}$ & 0.4971 & 0.6895 & 3.1750 & SSVB & 1.3871 \\ 
    			$x_{03}$ & 0.6124 & 0.8285 & 2.2023 & SSVB & 1.3528 \\ 
    			$x_{04}$ & 0.3072 & 0.4209 & 2.2970 & SSVB & 1.3700 \\ 
   				$x_{05}$ & 0.4469 & 0.6323 & 2.2751 & SSVB & 1.4149 \\ 
   				$x_{06}$ & 0.5602 & 0.7171 & 2.2786 & SSVB & 1.2800 \\ 
    			$x_{07}$ & 0.7639 & 0.8550 & 2.2397 & SSVB & 1.1193 \\ 
    			$x_{08}$ & 0.5345 & 0.7589 & 2.7780 & SSVB & 1.4197 \\ 
    			$x_{09}$ & 0.6695 & 0.6696 & 2.4548 & SSVB & 1.0001 \\ 
   				$x_{010}$ & 0.4239 & 0.6273 & 2.3603 & SSVB & 1.4797 \\ 
				\hline
		\end{tabular}}
	\end{subtable}
	\caption{Sample standard deviation for the 100 estimates in the Lorenz-96 model with 10 variables.}
	\label{tab:10LorenzStd}
\end{table}

\begin{figure}[t!]
	\centering
	\includegraphics[width=0.75\linewidth]{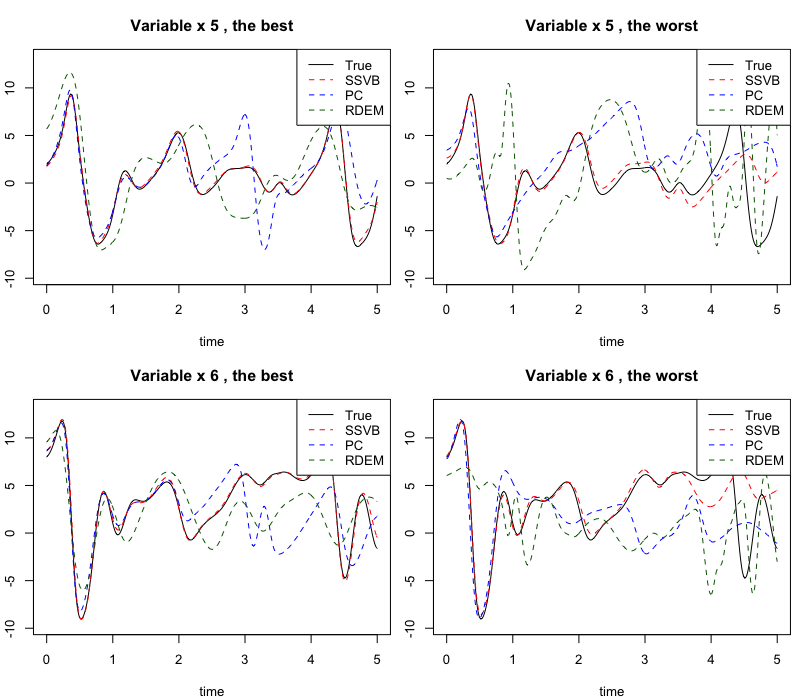}
	\caption{The estimated solution curves for $X_5$ and $X_6$ with their best/worst estimates ($\hat{\btheta}$, $\hat{\bfx}_0$) from the 100 data sets, of the Lorenz-96 model with 10 variables, for each method. The other variables ($X_1\sim X_4$ and $X_7\sim X_{10}$) also showed similar patterns. }
	\label{fig:10LorenzBestWorst}
\end{figure}

The most important result in this paper is shown in Figure~\ref{fig:10LorenzBestWorst}. 
Unlike the previous experiments, the best estimates of both PC and RDEM methods produced solution curves that significantly deviate from the true solution curve for all the 10 variables. This indicates that virtually all the 100 estimates from the 100 different data sets cannot provide valid regression curves. Of course, some of the 40 parameters could be estimated to be very close to the true parameters. However, solution curves of ODEs are  drastically changed by all the parameters in the ODE, including the initial states $\bfx_0$. In other words, if the estimates of the parameters do not form an appropriate combination, it is impossible to estimate the regression curve.

The strength of the SSVB is revealed here. Not only did the SSVB produce a solution curve that closely matches the true curve from the best estimate, but it also produced a regression curve that follows the true solution curve to some extent, even from the worst estimate. That is, in addition to estimating the parameters themselves properly, the solution curves derived from them were also consistent with the data without losing much of its stability even in the big model. 

The results are thought to be due to the concentrativeness of the mean-field variational method.  As the ODE system becomes more complex and larger, the ODE curve determined by $\bfx_{0}$ and  $\btheta$ can more sensitively vary in shape. That is, the combination of $\bfx_{0}$ and $\btheta$ becomes more important rather than the respective estimates. For the PC method, though the ODE parameters $\btheta$ are optimized under the interaction with the whole B-spline basis, the initial states $\bfx_{0}$, given $\btheta$, are only determined by some local basis coefficients of the several bases near $t=0$. For the MCMC based methods, including RDEM, as the model grows, the problems in moving on the parameter space also get bigger due to their complex dependencies between the parameters. On the contrary, the mean-field variational Bayes approximates the density to the one with dense density in the center. In other words, the assumption of independence in the mean-field makes the approximated density concentrated on the most representative combinations of $\bfx_{0}$ and $\btheta$. As a result, the reproduction of the ODE curves can be performed best through the proposed method.

\section{Application to real data: COVID-19} \label{s:appl}
\subsection{SIR model with time-varying parameters}
As an application to real-world data, we chose the COVID-19 epidemic, which is now the world’s biggest issue. Since the first outbreak in Wuhan, China, lots of observational data have been recorded daily. One important indicator is the number of infected people because it can be used to determine the infectivity of the disease. 

The SIR model of \citeasnoun{kermack1927contribution} using ODEs is one of the most famous tools for dealing with it. It has three compartments: S (Susceptible), I (Infectious), and R (Removed). S stands for the number of non-infected individuals who may be infected in the future.  I indicates the number of now infected people who can pass the infections to S. Lastly, R, meaning the removed people from the infectious relationships, is the number of recovered or dead people from the epidemic. They are not infectious and are no longer subject to infection because they have immunity or are already dead. The total population is divided into these three compartments.

Although there are diverse variants, the basic SIR model is as follows: 
\begin{align*}
\frac{dS(t)}{dt}&=-\frac{\beta I(t)S(t)}{N},\\
\frac{dI(t)}{dt}&=\ \frac{\beta I(t)S(t)}{N}-\gamma I(t), \\
\frac{dR(t)}{dt}&=\ \gamma I(t).
\end{align*}
The total population $N = S(t)+I(t)+R(t)$ is assumed to be constant. The parameter $\beta$ is the average number of contagious contacts by one infected person per unit time. 
Among a total of $\beta I(t)$  who had contagious contacts with infectious people, as much as the ratio $ S(t)/N $ of them can be newly infected and introduced into $I(t)$. On the other hand, an infected individual recovers or dies at an average rate of $\gamma$ and move on to the $R(t)$ compartment. In other words, it takes an average of $1/\gamma$ days to die or recover completely. It is known that the ratio of two parameters, the so-called basic reproductive number $R_0 = \beta/\gamma$, is used to evaluate the infectivity. 

Unfortunately, however, it is difficult to apply the above model to real-world data as it is. The model assumes that people’s contacts are simply random, such as those occurring in molecular motion. When a country with a population of tens of millions is taken as the unit of analysis, this assumption can lead to a significant gap from the real-world data. Actually, the forms of solutions that the model can have are particularly limited. For example, if you look at it briefly, it is easy to see that $R(t)$ has the steepest slope when $I(t)$ is at its maximum value since $\gamma$ is a constant. And when you check the actual data, you can see that this is rarely happening. The simple SIR model cannot fit the COVID-19 data.

Focusing on taking advantage of the SSVB algorithm at the same time as solving the above problem, we can think of the following model that gives the parameters great flexibility using the cubic B-spline basis:
\begin{align}
\label{TVSIR}
\begin{split}
\frac{dS(t)}{dt}&=-\frac{\beta(t) I(t)S(t)}{N},\\
\frac{dI(t)}{dt}&=\ \frac{\beta(t) I(t)S(t)}{N}-\gamma(t) I(t), \\ 
\frac{dR(t)}{dt}&=\ \gamma(t) I(t), \\ \noalign{\vskip10pt}
\text{where }\ \beta(t) &:= \exp\Big\{\sum_\ell c_{\beta,\ell}B_{\beta,\ell}(t)\Big\}, \\
\gamma(t) &:= \exp\Big\{\sum_\ell c_{\gamma,\ell} B_{\gamma,\ell}(t)\Big\}.
\end{split}
\end{align}
Here, $B_{\beta,\ell}(t)$ and $B_{\gamma, \ell}(t)$ are the cubic B-spline basis functions defined by the range of time $t$ and some equidistant knots. Since a B-spline curve is a piecewise polynomial function and is locally fitted to data, it can cope with well locally varying fluctuations. Especially in the case of order 4, which represents piecewise cubic polynomial, it has a continuous second derivative and forms a very smooth curve. The ODE parameters to estimate are now the coefficients $c_{\beta,\ell}$ and $c_{\gamma,\ell}$ of the basis functions, whose number can be increased by adding more knots. The exponential function was used to ensure positive values. 

\subsection{Fitting the COVID-19 data}
The SIR model looks as if it has three variables, but the actual variables are two since it assumes that the total population $N = S(t)+I(t)+R(t)$ is constant. Therefore, for practical implementation, we used the following equations with the substitution $ S(t)=N-I(t)-R(t)$:
\begin{align*}
\frac{dI(t)}{dt}&=\ \frac{\beta(t) I(t)(N-I(t)-R(t))}{N}-\gamma(t) I(t), \\ 
\frac{dR(t)}{dt}&=\ \gamma(t) I(t).
\end{align*}

The priors were:
\begin{align*}
\lambda\sim& \ \text{Gamma}(0.01,\ 0.01), \\
\theta_k\sim&\ \text{Unif }(-\infty,\ \infty),\ \text{ for all } k=1,\dots,q.
\end{align*} 
Since the scale of variable values is much larger than in the previous simulations, a flatter prior was given for $\lambda$. The number of ODE parameters, $q$, is determined by how many basis functions we use in (\ref{TVSIR}). We determined this based on the Bayesian information criterion (BIC) with the Bayes estimator. As a result of comparing models with 20, 25, 30, 35, and 40 basis functions, models with 30, 25, and 35 basis functions were selected for South Korea, China, and Japan data, respectively. In other words, the 60, 50, and 70 ODE parameters and the 2 initial states ​​were estimated for each country. The tuning parameters $m=1$, $\tau= 0.1^4$ were used.
\begin{figure}
	\centering
	\includegraphics[width=0.8\linewidth]{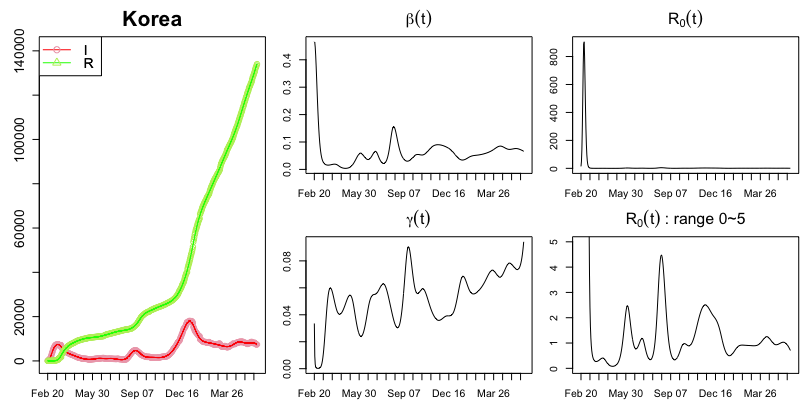}
	
	\vspace{3mm}
	\includegraphics[width=0.8\linewidth]{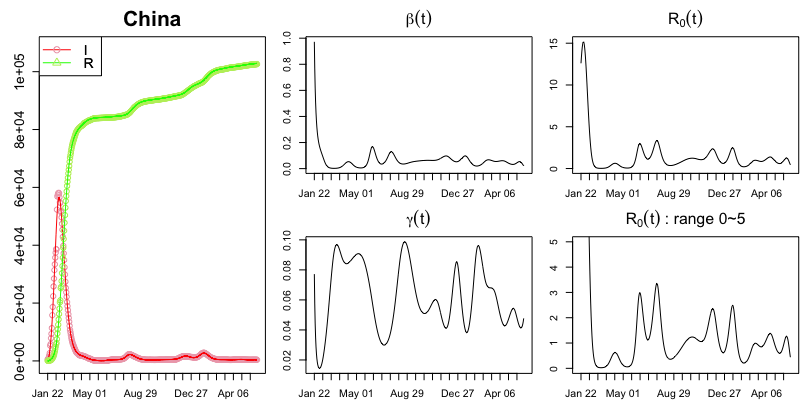}
	
	\vspace{3mm}
	\includegraphics[width=0.8\linewidth]{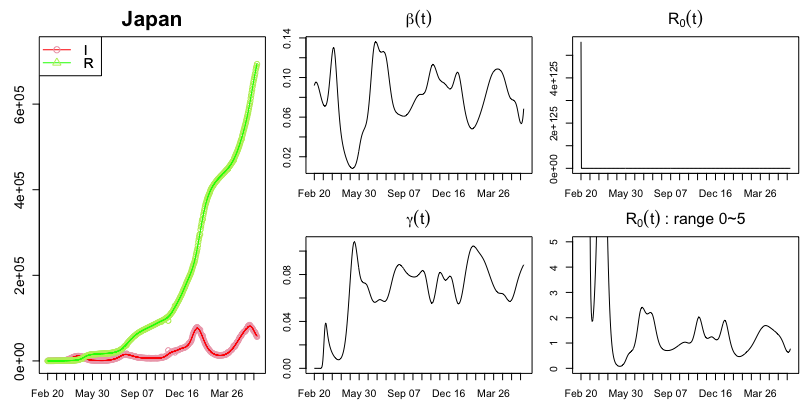}
	\caption{The fitted results for South Korea, China, and Japan. The four graphs on the right side show the estimated $\beta(t), \gamma(t)$, and $R_0(t)=\beta(t)/\gamma(t)$.}
	\label{fig:CovidResult}
\end{figure}

Because the variable values ​​are large and the ODE model in hand is sensitive to changes in parameters through exponential functions in $\beta(t)$ and $\gamma(t)$, the optimization process may fail to calculate gradients or take a long time to find an appropriate starting point. Accordingly, a good starting point calculated from the data in a discretized way was provided, and the learning rate of the line search in optimization was also given very small so that the algorithm could start successfully. The termination conditions were also given more loosely to prevent unnecessary ticking-over.

Figure~\ref{fig:CovidResult} shows the fitted results. The data on the number of confirmed cases in each country were obtained from the GitHub repository operated by the Center for Systems Science and Engineering (CSSE) at Johns Hopkins University, \citeasnoun**{dong2020interactive}. For the total population numbers $N$, we used the
UN World Population Prospects 2019 provided by \citeasnoun{UN2019}. The graph on the left shows the fitted results with the real data points, and the four graphs on the right show the estimated $\beta(t), \gamma(t)$, and $R_0(t)=\beta(t)/\gamma(t)$. For $R_0(t)$, graphs ranging from 0 to 5 were added to see the values ​​in a relatively realistic range. In the case of Korea, the data in the early days, which are so flat to violate the softness of the SIR model significantly, are not included in the inference. In the case of Japan, too, an excessively flat period in the early days was excluded from the estimation.

The SSVB algorithm properly estimated the ODE parameters within about $2\sim3$ minutes for each country.  As expected, the rapid fluctuations of beta and gamma were required to fit the COVID-19 data with the SIR model. This shows that the assumption of the simple SIR model is very unrealistic at the national scale and that much more sophisticated models are required for the analysis of epidemic spread patterns in the real world. It is meaningful that we directly verified this by estimating a number of parameters using the proposed algorithm.

\section{Discussion} \label{s:discuss}
We proposed an ODE parameter estimating method based on the state-space model and variational Bayes approximation. The conversion to the state-space model reduces the amount of computation from the complete numerical solution to a few one-step calculations at the observation time points. In addition, the posterior approximation by the variational Bayes makes the inference even faster. 

The proposed method showed strength not only in the speed but also in the performance of reproducing the ODE curves. In the simulation studies, it was especially noticeable as the number of parameters to be inferred increased. When the ODE model is large, the approximation of the variational Bayes has some advantage in providing a good combination of  $\hat{\bfx}_{0}$ and $\hat{\btheta}$. When applied to actual data with somewhat large variable values, it also showed decent performance in a few minutes.

Although the performance is good, underestimation of the posterior variance in the mean-field variational method remains to be improved. Several correction strategies have been studied, including the method by \citeasnoun**{giordano2015linear}, but due to the nature of the state-space model, which has a strong dependency between neighboring latent variables, it did not work well in our model. Nevertheless, those can be seen as a trade-off taken for better reproduction of the ODE curve. Considering the sensitivity of the ODE curves, the accuracy of estimation can be more valuable than the covariance structure.

\newpage
\appendix 
\section{Appendix: Optimization algorithm}\label{algorithm}
\begin{algorithm}
	\caption{Optimization algorithm of the SSVB}
	1. Update the mean parameters $\bfu = (\bmu,\ \bfm)$, when $ \bfm=(\bfm_0,\dots,\bfm_n)^T $.\\
	\hspace*{3mm} $ < $ approximate Riemannian conjugate gradient learning $ > $\\
	\hspace*{1cm} $\bfp_0 = \bzero_{q+p(n+1)} $, $\ \tilde{\bfg}_0 = \bone_{q+p(n+1)} $\\
	\hspace*{1cm} \textbf{for} $k=1,2,\dots$ \textbf{do} {\hfill $ \triangleright $ Repeat until convergence}\\
	\hspace*{2cm} $\tilde{\bfg}_k \leftarrow \text{diag}(\boldsymbol{\sigma}^2, \bfV)\nabla_{\bfu_{k-1}}C_{KL}(\bfu_{k-1},\ \cdot\ )${\hfill $\triangleright$ Riemannian gradient}\\
	\hspace*{2cm} $\beta \leftarrow \frac{(\nabla_{\bfu_{k-1}}C_{KL})^T (\tilde{\bfg}_k-\tilde{\bfg}_{k-1})}{\tilde{\bfg}_{k-1}^T\text{diag}\left(\frac{1}{\boldsymbol{\sigma}^2}, \frac{1}{\bfV}\right)\tilde{\bfg}_{k-1}}$ {\hfill $\triangleright$ Polak-Ribi\'{e}re formula}\\
	\hspace*{2cm} $\bfp_k \leftarrow -\tilde{\bfg}_k + \beta\bfp_{k-1}$ {\hfill $\triangleright$ Update direction}\\
	\hspace*{2cm} $\alpha \leftarrow \underset{\alpha}{\arg\min}\ C_{KL}(\bfu_{k-1} + \alpha\bfp_k,\ \cdot\ )$ {\hfill $ \triangleright $ Line search} \\
	\hspace*{2cm} $\bfu_k \leftarrow \bfu_{k-1} + \alpha\bfp_k$ {\hfill $ \triangleright $ Update} \\
	
	\noindent 2. Update the variance parameters $\bfs = (\bsigma^2, \bfV) $, when  $\bfV=(\bfV_0,\dots,\bfV_n)^T$ as vectors. \\
	\hspace*{3mm} $<$ fixed-point iteration $>$ \\
	\hspace*{1cm} $\bfs_{k}^{-1} \overset{iteration}{\Longleftarrow} 2\nabla_{\bfs_{k-1}}F_{\text{fixed}}(\bfs)$  \hfill $ \triangleright $ Iterate until convergence
	\begin{align*}
	F_{\text{fixed}}&(\bfs) \\ =&\ A_\lambda\log B_\lambda(\bfm,\bfV) +\frac{1}{2\tau}\sum_{i=1}^{n}\sum_{j=1}^{p}V_{ij}\\
	&+\frac{1}{2\tau}\frac{1}{M}\sum_{i=1}^{n}\sum_{s=1}^{M}\left\lVert\bfm_i-\bfg(\sqrt{\bfV_{i-1}}\odot\bfZ_{\bfx_{i-1}}^{(s)}+\bfm_{i-1},\ t_{i-1},\sqrt{\bsigma^2}\odot\bfZ_{\btheta}^{(s)}+\bmu)\right\rVert^2 
	\end{align*}
	
	\noindent 3. Iterate 1 and 2 until convergence.
\end{algorithm}

\bi
\item  We apply the \textit{quadratic interpolation method with three points} in \citeasnoun{sun2006optimization} for the line search procedure.  
\item As a skill to speed up the algorithm, we let the termination criteria of the early part of updating the mean parameters be lax and tighten it gradually up to the final criterion $\varepsilon$. It can improve the speed about 2 to 4 times faster.
\item The details for the computations are as follows. The notation $    \bfJ_{\bfg\text{ wrt }\btheta}(\cdot)   $ means the Jacobian matrix of $\bfg(\cdot)$ with respect to $\btheta$ and the others are analogous to it. 
\ei

\subsection{Gradient calculations for the Riemannian gradient}
The gradient vector with respect to the mean parameters consists of the sub-vectors:
\begin{equation*}
\begin{split}
\nabla_{\bfu}C_{KL} = 
\begin{bmatrix} 
\nabla_{\bmu}C_{KL} \\ 
\nabla_{\bfm_0}C_{KL} \\ 
\vdots \\
\nabla_{\bfm_n}C_{KL}  
\end{bmatrix}.
\end{split}
\end{equation*}
Using matrix calculus and the chain rule, we can obtain the first part,
\begin{equation*}
\begin{split}
\nabla_{\bmu}C_{KL} =& \nabla_{\bmu} \left[\frac{1}{2\tau}\sum_{i=1}^{n}\frac{1}{M}\sum_{s=1}^{M}\left\lVert\bfm_i-\bfg(\bfx_{i-1}^{(s)},t_{i-1},\sqrt{\bsigma^2}\odot\bfZ_{\btheta}^{(s)}+\bmu)\right\rVert^2\right] \\
=& -\frac{1}{\tau}\frac{1}{M}\sum_{i=1}^{n}\sum_{s=1}^{M}\bfJ_{\bfg\text{ wrt }\btheta}\left(\bfx_{i-1}^{(s)},t_{i-1},\btheta^{(s)}\right)^T \left(\bfm_i-\bfg(\bfx_{i-1}^{(s)},t_{i-1},\btheta^{(s)})\right).
\end{split}
\end{equation*}
In the same way,
\begin{equation*}
\begin{split}
\nabla_{\bfm_0}C_{KL} =&\ \frac{A_\lambda}{B_\lambda(\bfm,\bfV)}(\bfm_0-\bfy_0) \\
& -\frac{1}{\tau}\frac{1}{M}\sum_{s=1}^{M}\bfJ_{\bfg\text{ wrt }\bfx}\left(\bfx_{0}^{(s)},t_{0},\btheta^{(s)}\right)^T \left(\bfm_1-\bfg(\bfx_{0}^{(s)},t_{0},\btheta^{(s)})\right),
\end{split}
\end{equation*}
for $i=1,\cdots, n-1$,
\begin{equation*}
\begin{split}
\nabla_{\bfm_i}C_{KL} =&\ \frac{A_\lambda}{B_\lambda(\bfm,\bfV)}(\bfm_i-\bfy_i)  + \frac{1}{\tau}\frac{1}{M}\sum_{s=1}^{M} \left(\bfm_i-\bfg(\bfx_{i-1}^{(s)},t_{i-1},\btheta^{(s)})\right)\\
& -\frac{1}{\tau}\frac{1}{M}\sum_{s=1}^{M}\bfJ_{\bfg\text{ wrt }\bfx}\left(\bfx_{i}^{(s)},t_{i},\btheta^{(s)}\right)^T \left(\bfm_{i+1}-\bfg(\bfx_{i}^{(s)},t_{i},\btheta^{(s)})\right),
\end{split}
\end{equation*}

\begin{equation*}
\begin{split}
\nabla_{\bfm_n}C_{KL} =&\ \frac{A_\lambda}{B_\lambda(\bfm,\bfV)}(\bfm_n-\bfy_n)  + \frac{1}{\tau}\frac{1}{M}\sum_{s=1}^{M} \left(\bfm_n-\bfg(\bfx_{n-1}^{(s)},t_{n-1},\btheta^{(s)})\right).
\end{split}
\end{equation*}
\subsection{Gradient calculations for the fixed-point iteration}
For the fixed-point iteration,
\begin{equation*}
\begin{split}
2\nabla_{\bfs}F_{\text{fixed}}(\bfs) = 
\begin{bmatrix} 
2\nabla_{\bsigma^2}F_{\text{fixed}}(\bfs)\\ 
2\nabla_{\bfV_0}F_{\text{fixed}}(\bfs) \\ 
\vdots \\
2\nabla_{\bfV_n}F_{\text{fixed}}(\bfs)
\end{bmatrix}.
\end{split}
\end{equation*}
In the same way as above, we use matrix calculus and the chain rule.
\begin{equation*}
\begin{split}
2\nabla_{\bsigma^2}F_{\text{fixed}}(\bfs) =& -\frac{1}{\tau}\frac{1}{M}\sum_{i=1}^{n}\sum_{s=1}^{M}\text{diag}\left[\frac{1}{\bsigma}\odot\bfZ_{\btheta}^{(s)}\right]\bfJ_{\bfg\text{ wrt }\btheta}\left(\bfx_{i-1}^{(s)},t_{i-1},\btheta^{(s)}\right)^T \left(\bfm_i-\bfg(\bfx_{i-1}^{(s)},t_{i-1},\btheta^{(s)})\right),\\
2\nabla_{\bfV_0}F_{\text{fixed}}(\bfs) =& \ \frac{A_\lambda}{B_\lambda(\bfm,\bfV)}\bfone_p \\ &-\frac{1}{\tau}\frac{1}{M}\sum_{s=1}^{M}\text{diag}\left[\frac{1}{\sqrt{\bfV_0}}\odot\bfZ_{\bfx_0}^{(s)}\right]\bfJ_{\bfg\text{ wrt }\bfx}\left(\bfx_{0}^{(s)},t_{0},\btheta^{(s)}\right)^T \left(\bfm_1-\bfg(\bfx_{0}^{(s)},t_{0},\btheta^{(s)})\right),
\end{split}
\end{equation*}
for $i=1,\cdots, n-1$,
\begin{equation*}
\begin{split}
2\nabla_{\bfV_i}F_{\text{fixed}}(\bfs) =& \ \frac{A_\lambda}{B_\lambda(\bfm,\bfV)}\bfone_p +\frac{1}{\tau}\bfone_p \\ &-\frac{1}{\tau}\frac{1}{M}\sum_{s=1}^{M}\text{diag}\left[\frac{1}{\sqrt{\bfV_i}}\odot\bfZ_{\bfx_i}^{(s)}\right]\bfJ_{\bfg\text{ wrt }\bfx}\left(\bfx_{i}^{(s)},t_{i},\btheta^{(s)}\right)^T \left(\bfm_{i+1}-\bfg(\bfx_{i}^{(s)},t_{i},\btheta^{(s)})\right),\\
2\nabla_{\bfV_n}F_{\text{fixed}}(\bfs) =& \ \frac{A_\lambda}{B_\lambda(\bfm,\bfV)}\bfone_p +\frac{1}{\tau}\bfone_p .
\end{split}
\end{equation*}
\\
The calculations of Jacobian matrices in the formulas above are derived in Appendix \ref{Jacobian}.

\section{Appendix: Jacobian matrix}\label{Jacobian}
The proposed method needs the Jacobian matrix of the approximating function $\bfg(\cdot)$ in (\ref{SSM}). We present the derivation of it from that of the ODE function $\bff(\cdot)$, $ \bfJ_{\bff} $, in the case of the 4th-order Runge-Kutta method: 
\begin{align}
\begin{split}
\bfg(\bfx,t,\btheta)=&\ \bfx+\frac{1}{6}(K_{1}+2K_{2}+2K_{3}+K_{4}),\\
K_{1}=&\ h\cdot \bff(\bfx,t;\btheta),\\
K_{2}=&\ h\cdot \bff\left(\bfx+\frac{1}{2}K_{1},t+\frac{1}{2}h;\btheta\right),\\
K_{3}=&\ h\cdot \bff\left(\bfx+\frac{1}{2}K_{2},t+\frac{1}{2}h;\btheta\right),\\
K_{4}=&\ h\cdot \bff(\bfx+K_{3},t+h;\btheta),
\end{split}
\end{align}
which is selected in this paper. For simplicity, the subscripts are omitted, and equally spaced observation times ($h = h_1 = \cdots = h_n $) are assumed.

\subsection{The case of step size $m=1$}\label{Jacobian_m1}
First, when the step size $m=1$, the Jacobian of $\bfg(\cdot)$ with respect to $\bfx$ can be obtained by 
$$ \bfJ_{\bfg\text{ wrt }\bfx}(\bfx,t,\btheta)=\ \bfI_{p\times p} +\frac{1}{6}(\bfJ_{K_1\text{ wrt }\bfx}+2\bfJ_{K_2\text{ wrt }\bfx}+2\bfJ_{K_3\text{ wrt }\bfx}+\bfJ_{K_4\text{ wrt }\bfx}),$$
\begin{align*}
\begin{split}
\bfJ_{K_1\text{ wrt }\bfx}=&\ h\cdot \bfJ_{\bff\text{ wrt }\bfx}(\bfx, t,\btheta),\\
\bfJ_{K_2\text{ wrt }\bfx}=&\ h\cdot \bfJ_{\bff\text{ wrt }\bfx}\left(\bfx+\frac{1}{2}K_1, t+\frac{1}{2}h,\btheta\right)
\begin{bmatrix} 
\bfI_{p\times p} + \frac{1}{2}\bfJ_{K_1\text{ wrt }\bfx}
\end{bmatrix},\\
\bfJ_{K_3\text{ wrt }\bfx}=&\ h\cdot \bfJ_{\bff\text{ wrt }\bfx}\left(\bfx+\frac{1}{2}K_2, t+\frac{1}{2}h,\btheta\right)
\begin{bmatrix} 
\bfI_{p\times p} + \frac{1}{2}\bfJ_{K_2\text{ wrt }\bfx}
\end{bmatrix},\\
\bfJ_{K_4\text{ wrt }\bfx}=&\ h\cdot \bfJ_{\bff\text{ wrt }\bfx}(\bfx+K_3, t+h,\btheta)
\begin{bmatrix} 
\bfI_{p\times p} + \bfJ_{K_3\text{ wrt }\bfx}
\end{bmatrix},\\
\end{split}
\end{align*}
from using the matrix chain rule.

On the other hand, with respect to $\theta$, we should be cautious that $K_2,\ K_3$, and $ K_4 $ include the previous term that is also a function of $\btheta$, in the position of $\bfx$ argument. Finally, we get:
$$ \bfJ_{\bfg\text{ wrt }\btheta}(\bfx,t,\btheta)=\ \frac{1}{6}(\bfJ_{K_1\text{ wrt }\btheta}+2\bfJ_{K_2\text{ wrt }\btheta}+2\bfJ_{K_3\text{ wrt }\btheta}+\bfJ_{K_4\text{ wrt }\btheta}), $$
\begin{align*}
\begin{split}
\bfJ_{K_1\text{ wrt }\btheta}=&\ h\cdot \bfJ_{\bff\text{ wrt }\btheta}(\bfx, t,\btheta),\\
\bfJ_{K_2\text{ wrt }\btheta}=&\ h\cdot
\begin{bmatrix}
\bfJ_{\bff\text{ wrt }\bfx}\left(\bfx+\frac{1}{2}K_1, t+\frac{1}{2}h,\btheta\right)\quad \bfJ_{\bff\text{ wrt }\btheta}\left(\bfx+\frac{1}{2}K_1, t+\frac{1}{2}h,\btheta\right)
\end{bmatrix}
\begin{bmatrix} 
\half\bfJ_{K_1\text{ wrt }\btheta}\\ 
\bfI_{q\times q}
\end{bmatrix}, \\
\bfJ_{K_3\text{ wrt }\btheta}=&\ h\cdot
\begin{bmatrix}
\bfJ_{\bff\text{ wrt }\bfx}\left(\bfx+\frac{1}{2}K_2, t+\frac{1}{2}h,\btheta\right)\quad \bfJ_{\bff\text{ wrt }\btheta}\left(\bfx+\frac{1}{2}K_2, t+\frac{1}{2}h,\btheta\right)
\end{bmatrix}
\begin{bmatrix} 
\half\bfJ_{K_2\text{ wrt }\btheta}\\ 
\bfI_{q\times q}
\end{bmatrix}, \\
\bfJ_{K_4\text{ wrt }\btheta}=&\ h\cdot
\begin{bmatrix}
\bfJ_{\bff\text{ wrt }\bfx}\left(\bfx+K_3, t+h,\btheta\right)\quad \bfJ_{\bff\text{ wrt }\btheta}\left(\bfx+K_3, t+h,\btheta\right)
\end{bmatrix}
\begin{bmatrix} 
\bfJ_{K_3\text{ wrt }\btheta}\\ 
\bfI_{q\times q}
\end{bmatrix}.
\end{split}
\end{align*}
These Jacobian matrices (functions) become the base functions of the following cases. 

\subsection{The case of step size $m\geq2$}
First, without considering the step size $m$, we can define some iterative functions like:
\begin{align*}
\bfg^{(2)}(\bfx,t,\btheta)=&\ \bfg(\bfg(\bfx,t,\btheta),t+h,\btheta),\\
\bfg^{(3)}(\bfx,t,\btheta)=&\ \bfg(\bfg^{(2)}(\bfx,t,\btheta),t+2h,\btheta),\\ &\vdots \\
\bfg^{(m)}(\bfx,t,\btheta)=&\ \bfg\left(\bfg^{(m-1)}(\bfx,t,\btheta),t+(m-1)h,\btheta\right).	 
\end{align*}

Then, the Jacobian matrices with respect to $\bfx$ of the above functions can be recursively computed from $\bfJ_{\bfg\text{ wrt }\bfx}(\cdot)$ of \ref{Jacobian_m1} as follows: 
\begin{align*}
\bfJ_{\bfg^{(2)}\text{ wrt }\bfx}(\bfx,t,\btheta) =&\ \bfJ_{\bfg\text{ wrt }\bfx}\left(\bfg(\bfx,t,\btheta),\ t+h,\ \btheta\right)\bfJ_{\bfg\text{ wrt }\bfx}(\bfx,t,\btheta),\\
\bfJ_{\bfg^{(3)}\text{ wrt }\bfx}(\bfx,t,\btheta) =&\ \bfJ_{\bfg\text{ wrt }\bfx}\left(\bfg^{(2)}(\bfx,t,\btheta),\ t+2h,\ \btheta\right)\bfJ_{\bfg^{(2)}\text{ wrt }\bfx}(\bfx,t,\btheta),\\
&\vdots\\
\bfJ_{\bfg^{(m)}\text{ wrt }\bfx}(\bfx,t,\btheta) =&\ \bfJ_{\bfg\text{ wrt }\bfx}\left(\bfg^{(m-1)}(\bfx,t,\btheta),\ t+(m-1)h,\ \btheta\right)\bfJ_{\bfg^{(m-1)}\text{ wrt }\bfx}(\bfx,t,\btheta)
.	 
\end{align*}

For $\btheta$, again, be cautious that each function includes the previous function which is a function of $\btheta$, in the position of $\bfx$ argument. The chain rule of matrix calculus gives:
\begin{align*}
\bfJ_{\bfg^{(2)}\text{ wrt }\btheta}(\bfx,t,\btheta) &=
\begin{bmatrix}
\bfJ_{\bfg\text{ wrt }\bfx}\left(\bfg(\bfx,t,\btheta),\ t+h,\ \btheta\right)\quad
\bfJ_{\bfg\text{ wrt }\btheta}\left(\bfg(\bfx,t,\btheta),\ t+h,\ \btheta\right)
\end{bmatrix}
\begin{bmatrix}
\bfJ_{\bfg\text{ wrt }\btheta}(\bfx,t,\btheta) \\
\bfI_{q\times q}
\end{bmatrix},\\
\bfJ_{\bfg^{(3)}\text{ wrt }\btheta}(\bfx,t,\btheta) &=
\begin{bmatrix}
\bfJ_{\bfg\text{ wrt }\bfx}\left(\bfg^{(2)}(\bfx,t,\btheta),t+2h,\btheta\right)\quad
\bfJ_{\bfg\text{ wrt }\btheta}\left(\bfg^{(2)}(\bfx,t,\btheta),t+2h,\btheta\right)
\end{bmatrix}
\begin{bmatrix}
\bfJ_{\bfg^{(2)}\text{ wrt }\btheta}(\bfx,t,\btheta) \\
\bfI_{q\times q}
\end{bmatrix},\\
&\quad\vdots \\
\bfJ_{\bfg^{(m)}\text{ wrt }\btheta}(\bfx,t,\btheta)& \\
& \hspace{-3cm}
={\small \begin{bmatrix}
	\bfJ_{\bfg\text{ wrt }\bfx}\left(\bfg^{(m-1)}(\bfx,t,\btheta),t+(m-1)h,\btheta\right)\quad
	\bfJ_{\bfg\text{ wrt }\btheta}\left(\bfg^{(m-1)}(\bfx,t,\btheta),t+(m-1)h,\btheta\right)
	\end{bmatrix}
	\begin{bmatrix}
	\bfJ_{\bfg^{(m-1)}\text{ wrt }\btheta}(\bfx,t,\btheta) \\
	\bfI_{q\times q}
	\end{bmatrix}}.
\end{align*}

Now to return to the main, for the given step size $m\geq2$ in the proposed method, we can use $\bfJ_{\bfg^{(m)}\text{ wrt }\bfx}(\bfx,t,\btheta)$ and $ \bfJ_{\bfg^{(m)}\text{ wrt }\btheta}(\bfx,t,\btheta) $ that are computed with $h/m$ instead of $h$ from the above formulas as the Jacobian matrix of $\bfg(\cdot)$ in the algorithm of Appendix \ref{algorithm} .

\section{Appendix: How to determine the tuning parameters}\label{tuning}
The tuning parameters $m$ and $\tau$ can be important to fit the state-space model to the data from the original ODE model. For example, if the step size $m=1$ and the observation time interval is wide, it would be appropriate not to make $\tau$ too small. Because the next state approximated using the function $\bfg$ and the one integrated using a (solution) solver may differ considerably. The information available to determine $m$ and $\tau$ in a given data set is as follows:
\begin{itemize}
	\item Observation time points in the data, $ t_0\leq t_1\leq\cdots\leq t_n. $
	\item Roughly possible ranges for $\btheta$ (that is, the uniform prior $\pi(\btheta)$).
	\item The optimization starting point for $\bfx_0$ obtained using the (simple) cubic B-spline regression. Denote the starting point as $\bfx_0^*$.
\end{itemize}
\begin{algorithm}[hb!]
	\label{al:tau}
\caption{Reasonable $\tau$ for the given step size $m$.}
\begin{enumerate}
	\item Obtain a sample variance: 	\\
	\hspace*{0.5cm} Draw $\btheta^* \sim \pi(\btheta)$ and let $ \bfs^*(t) := \bfs(\bfx_0^*,t,\btheta^*) $.  \\
	\hspace*{0.5cm} Here, $ \bfs(\bfx_0^*,t,\btheta^*) $ is the solution curves generarated from $(\bfx_0^*,\btheta^*).$ \\
	\hspace*{0.5cm} \textbf{if}  $\ \ \underline{\bfY} \leq \bfs^*(t) \leq \overline{\bfY}\ \ $ for all $t_0\leq t\leq t_n$  when
	\begin{align*}
	&\bigg\{
	\begin{array}{ll}
	\overline{\bfY} & :=\max \bfY + 3\cdot(\max\bfY - \min\bfY)\\
	\underline{\bfY}& :=\min \bfY - 3\cdot(\max\bfY - \min\bfY),
	\end{array}
	\end{align*}
	\hspace*{0.5cm} \textbf{then} calculate the sample variance of $ \{d_{ij}|i = 1,\dots,n \text{ and }  j=1,\dots,p\} $ when
	$$ \bfd_i:= \bfg^{(m)}(\bfs^*(t_{i-1}),t_{i-1},\btheta^*)-\bfs^*(t_i) .$$
	
	\item Repeat 1  and obtain 100 sample variances.
	
	\item List the 100 sample variances ​​in order of size, and take the average value of the middle 50 (from 25th to 75th) values.  ​​Round up to one decimal point above the first non-zero value (ex. $0.000389 \ra  0.001$) and regard it as the reasonable $\tau$. 
\end{enumerate}
\end{algorithm}

Given the above information, $m$ and $\tau$ can be selected by the following procedure:
\begin{enumerate}
	\item For $m=1$, obtain a reasonable $\tau$ using Algorithm 2.
	\item If the obtained $\tau$ is greater than 0.0001, increase $m$ by 1 and calculate the reasonable $\tau$ using Algorithm 2 again.
	\item Repeat 2 until $\tau$ is not more than 0.0001 to obtain the final $m$ and $\tau$.
\end{enumerate}
Here, the maximum allowable value for fidelity to the ODE, 0.0001, was determined empirically.

The above procedure is only one recommendation, not absolute. Actually, the simulation experiments were conducted with the values ​​recommended by the above procedure, but for the COVID-19 real data fitting in Section \ref{s:appl}, $m=1$ was fixed considering the complexity of the time-varying SIR model. In this case, $\tau$ derived from Algorithm 2 was greater than 0.0001, but we selected $\tau=0.0001$, the maximum allowable value. Nevertheless, it showed good fitted results. In other words, you can try first with some values, such as $m=1$ and $\tau=0.0001$, and then adjust based on them.

\bibliographystyle{dcu}
\newpage
\bibliography{odevm}

\end{document}